\documentclass[12pt,english]{elsarticle}
\pdfoutput=1
\usepackage[T1]{fontenc}
\usepackage[latin9]{inputenc}
\usepackage{color}
\usepackage{float}
\usepackage{booktabs}
\usepackage{multirow}
\usepackage{varwidth}
\usepackage{amstext}
\usepackage{graphicx}

\makeatletter

\newcommand{\lyxmathsym}[1]{\ifmmode\begingroup\def\b@ld{bold}
  \text{\ifx\math@version\b@ld\bfseries\fi#1}\endgroup\else#1\fi}

\providecommand{\tabularnewline}{\\}
\newenvironment{cellvarwidth}[1][t]
    {\begin{varwidth}[#1]{\linewidth}}
    {\@finalstrut\@arstrutbox\end{varwidth}}
\newcommand{\lyxdot}{.}

\@ifundefined{date}{}{\date{}}
\usepackage{braket}

\usepackage{babel}

\makeatother

\usepackage{babel}
\begin{document}
\begin{frontmatter}
\title{Computational study of geometry, electronic structure, and low-lying
excited states of linear T-graphene quantum dots}
\author{Arifa Nazir}
\ead{arifabhatt9@gmail.com}
\address{Department of Physics, Indian Institute of Technology Bombay, Powai,
Mumbai 400076 India}
\author{Alok Shukla{*}}
\ead{{*}shukla@iitb.ac.in}
\address{Department of Physics, Indian Institute of Technology Bombay, Powai,
Mumbai 400076 India}
\begin{abstract}
A few years ago, by means of first-principles calculations, Enyashin
\emph{et al.} {[}Phys. Stat. Solidi B \textbf{248}, 1879 (2011){]}
proposed several novel monolayers of carbon containing rings other
than hexagons. One of those monolayers containing tetragons and octagons
was investigated later in detail by Liu \emph{et al.} {[}Phys. Rev.
Lett. \textbf{108}, 225505 (2012){]} who called it T-graphene, and
found that it exists both in strictly planar and buckled forms, with
the planar structure being metallic in nature. Given the fact\textcolor{magenta}{{}
}that Kotakoski \emph{et al.} {[}Phys. Rev. Lett. \textbf{106}, 105505
(2011){]} had already found experimental evidence of 1D carbon structures
containing tetragons and octagons, we decided to investigate finite
linear fragments of T-graphene, with the strictly planar structures,
referred to as T-graphene quantum dots (TQDs). In order to avoid the
dangling bonds in the finite T-graphene fragments, we considered the
edges to be saturated by hydrogen atoms. We first optimized the geometries
of the considered TQDs using a first-principles density-functional
theory (DFT) methodology, followed by calculations of their linear
optical absorption spectra using the time-dependent DFT (TDDFT) approach.
Given the fact that strictly planar T-graphene structures will have
$\sigma-\pi$ separation with the $\pi$ electrons near the Fermi
level, we also parameterized an effective $\pi$-electron Hamiltonian
for TQDs, similar to the Pariser-Parr-Pople model for $\pi$-conjugated
molecules. We further used the effective Hamiltonian to perform high-order
electron-correlated calculations using the configuration-interaction
(CI) approach to compute the optical absorption spectra of TQDs, and
also their singlet-triplet gaps. By considering the symmetries and
transition dipoles of the low-lying excited states, we found that
TQDs are photoluminescent materials. Moreover, in all the TQDs HOMO-LUMO
transition is optically forbidden, the optical gaps of these molecules
are quite large, suggesting the intriguing possibility of the fission
of a singlet optical exciton into several triplet excitons. 
\end{abstract}
\end{frontmatter}

\section{Introduction }

In the quest for novel materials with tailored electronic and optical
properties, carbon-based $\pi$-electron materials have emerged as
a fascinating class that bridges the worlds of chemistry, physics,
and materials science. These materials, characterized by their extended
$\pi$-electron systems derived from conjugated carbon networks, offer
a rich platform for exploring a wide range of electronic, optical,
and functional behaviors. With carbon's unique ability to form diverse
structures and arrangements, the manipulations of $\pi$-electrons
within these materials have paved the way for groundbreaking discoveries
and technological advancements. One of the most celebrated form (allotrope)
of carbon is graphene\citep{doi:10.1126/science.1102896}, which is
a monolayer of graphite with carbon atoms arranged in a hexagonal
fashion, giving rise to linear dispersion and Dirac cones in its electronic
band structure\citep{RevModPhys.81.109,novoselov2005two}. Due to
the presence of Dirac-like Fermions, graphene has amazing transport
propertie\textcolor{black}{s \citep{Wessely2014}, and has found numerous
applications in the field of nanoelectronics. The linear dispersion
characteristics in the electronic structure of graphene are generally
attributed to its honeycomb lattice. However, out of scientific curiosity
one wonders whether it is possible to have carbon monolayers without
the hexagonal rings? A few years ago, Enyashin }\textcolor{black}{\emph{et
al}}\textcolor{black}{.\citep{Enyashin2011} proposed twelve two-dimensional
non-hexagonal carbon allotropes. Later on, Liu }\textcolor{black}{\emph{et
al}}\textcolor{black}{.\citep{Liu2012} studied one of the proposed
structures, consisting of tetragons and octagons, in detail, and found
that it exists both in strictly planar and buckled forms. They found
that the planar one is metallic in nature, while the buckled monolayer
is a zero-gap semimetal exhibiting Dirac cones, just like graphene.
They named these structures as T-graphene because both possess a tetragonal
unit cell consisting of four carbon atoms.} In terms of stability,
both were found to be dynamically stable, with the buckled T-graphene
being more stable compared to other allotropes above 940 K, while
its planar counterpart is comparatively more stable (except graphene)
below 900 K. Sheng \textit{et al.} in their theoretical study on T-graphene
revealed that with the help of doping such as by boron-nitrogen pair,
its band gap opens up which results in a transition from semi-metallic
to semiconducting behavior\citep{Sheng_2012}, making it useful for
nanoelectronic applications. As far as the transport properties are
concerned, computational studies reveal that the presence of non-hexagonal
rings in 2D carbon allotropes including T-graphene, leads to a much
higher electronic component of the thermal conductivity as compared
to graphene\citep{Tong}. One can further tune the properties of T-graphene
by considering multilayer structures. For example, Bhattacharya and
Jana demonstrated by computational studies the stability of twin T-graphene,
an allotrope in which two tetragonal rings are placed one over the
other, resulting in a buckled three-atom thick layer of T-graphene,
exhibiting semiconducting nature, with an indirect band gap\citep{D0CP00263A}.

Planar T-graphene and its one-dimensional (1D) structures, namely
T-graphene nanoribbons (TGNRs), were studied by Wang \emph{et al.}\citep{C2CP41464C}
who found that: (a) armchair TGNR oscillates between metallic and
semiconducting behavior as a function of the width, while (b) zigzag
TGNR remains metallic regardless of the width. Electronic and transport
properties of TGNRs were studied by Dai \emph{et al}.\citep{Dai_2014}
employing the tight-binding model, as well as supported by the first-principles
calculations.\textcolor{black}{{} They found the armchair T-graphene
nanoribbons to have negative differential resistance, but zigzag T-graphene
nanoribbons show linear current-bias voltage characteristics close
to the Fermi level }\citep{Dai_2014}\textcolor{black}{. At further
reduced dimensions, electronic and optical properties of T graphene
quantum dots }(TQDs)\textcolor{black}{{} of varying sizes and shapes
were studied by Deb }\textcolor{black}{\emph{et al.}}\textcolor{black}{\citep{Deb2020}
using the first-principles approach, with their main focus on the
non-linear optical properties of the planar TQDs. It was revealed
that TQDs are energetically and dynamically stable and majority of
them show a strong non-linear optical response \citep{Deb2020}. }There
are several studies performed on T graphene from its application perspective
and it was found that T-graphene has potential uses in many different
fields, including optoelectronics \citep{C7CP03983B}, spintronics\citep{CHOWDHURY2017523},
current rectification\citep{article-bandyo}, hydrogen storage \citep{Sheng_2012},
and gas sensing\citep{article-bandyo,article-liu}. A study by Zhang
\textit{et al}. shows that T-graphene has a potential to serve as
an electrode for lithium and sodium ion batteries with a storage capacity
much higher compared to the graphite anodes available commercially\citep{ZHANG2020146849}.

\textcolor{black}{The field of carbon monolayers consisting of tetra
rings has recently been reviewed by Bandyopadhyay and Jana\citep{Bandyopadhyay_2020_review}. }

As far as the experimental realization is concerned, during the synthesis
of graphene, non-hexagonal ring configurations may spontaneously develop\citep{Lahiri2010}
as well as can be created deliberately by electron irradiation\citep{PhysRevLett.106.105505}.
A planar carbon monolayer called biphenylene which consists of tetragons,
hexagons, and octagons was recently synthesized in the laboratory
by Fan et al.\citep{biphynelene-exp}. Additionally, Kotakoski et
al.{} \citep{PhysRevLett.106.105505} in their measurements on amorphous
carbon monolayers, observed 1D structures consisting of carbon octagons
and tetrarings similar to TGNRs. Liu \textit{et al}. \citep{Liu2017_expt}\textit{
}were able to experimentally obtain 1D non-hexagonal nanoribbons of
carbon by embedding squares and octagons in graphitic nanostructures.
Their study revealed that incorporating four and eight-membered carbon
rings, enables precise control over the band gaps of carbon nanoribbons
and suppresses the spin-polarized edge states present in zigzag graphene
nanoribbons (GNRs). Thus, carbon allotropes with non-hexagonal carbon
rings offer a promising strategy for tailoring various properties
of carbon-based nanostructures to meet specific functional requirements.

\textcolor{black}{Motivated by}\textcolor{red}{{} }these findings\textcolor{red}{{}
}\textcolor{black}{we decided to explore the structure, stability,
and electro-optical properties of finite-sized linear fragments of
T-graphene, similar to the narrowest armchair-type TGNRs, referred
to as T-graphene quantum dots (TQDs) in this work. First we optimize
the geometry of each TQD using a first-principles density-functional
theory (DFT) based approach, with the dangling bonds of the edge carbon
atoms passivated by hydrogen atoms. Next, we compute the linear optical
absorption spectra of the considered TQDs employing the time-dependent
DFT (TDDFT) methodology. Although, the TDDFT approach is quite powerful,
however, it becomes time consuming with increasing size of the system.
Therefore, we use the information related to the excited states obtained
from the TDDFT calculations to construct an effective $\pi$-electron
Hamiltonian for the TQDs, with a structure similar to that of the
Pariser-Parr-Pople model Hamiltonian\citep{Pople1953,pariser1953-ppp}
used in chemistry to describe the electronic structure and optical
properties of $\pi$-conjugated molecules. With just one electron
and one $\pi$ orbital per carbon atom, the PPP model reduces the
electronic degrees of freedom of the system tremendously, thereby
allowing bigger electron-correlated calculations on larger-sized $\pi$-conjugated
molecules\citep{Gundra2013}. In our group, we have used PPP-model
based configuration-interaction (CI) approach to study a variety of
$\pi$-conjugated systems such as} conjugated polymers\citep{shukla2002correlated,PhysRevB.69.165218,sony2007large},
polycyclic aromatic hydrocarbons\citep{aryanpour2014subgap,pritam-ppp},
graphene nanoribbons\citep{gundra2011band,gundra2011theory}, and
graphene quantum dots\citep{tista-stone-wales}. We perform large-scale
CI calculations on the TQDs, within the framework of the PPP model,
to probe the many-electron nature of their optically excited states.
Furthermore, we also compute the spin gap of the TQDs using the same
PPP-CI methodology.

The remainder of this paper is as follows: in section \ref{sec:Computational-Approach}
we describe the basics of our computational approach and also the
parameterization of the PPP model for TQDs, next in section \ref{sec:Results-and-discussions}
we present and discuss our computational results, and finally in section
\ref{sec:Conclusion} we present our conclusions.

\section{Computational Approach}

\label{sec:Computational-Approach}

In this section, we briefly discuss the first-principles DFT based
approach used to perform the geometry optimization TQDs, and the calculation
of their optical absorption spectra using the TDDFT method. This is
followed by the details about the parameterization of the PPP model
applicable to TQDs using the TDDFT results, and then its subsequent
use within a CI approach to perform calculations of their absorption
spectra and the spin gaps.

\subsection{First-principles approach}

All the first-principles DFT calculations on the considered TQDs were
performed using the Gaussian16 \citep{Frisch2016} computational package,
while the visualization was done using the Gaussview package \citep{gv6}
and\textcolor{magenta}{{} }Xcrysden \citep{kokalj1999xcrysden}.
The geometry optimization of these structures was performed within
the DFT methodology using a triple-valence zeta basis set $6311$+g(d,p)\citep{krishnan1980self}
which includes both diffuse and polarization functions, coupled with
a hybrid exchange-correlation functional B3LYP\citep{1993JChPh..98.5648B,lee1988development,becke1992density}.
In order to avoid the dangling bonds, we saturated the edges of the
TQDs with hydrogen atoms. A self-consistent solution to the Kohn-Sham
equations\citep{kohn1965self} was obtained using a convergence threshold
of 10\textsuperscript{-}\textsuperscript{8} Hartree. The molecular
geometry was considered converged only when the maximum atomic force,
root-mean-square (RMS) force, maximum atomic displacement, and RMS
displacement were below 0.00045 Hartree/Bohr, 0.00030 Hartree/Bohr,
0.0018 Bohr, and 0.0012 Bohr, respectively.

For the first-principles TDDFT calculations used for computing the
optical absorption spectra of TQDs, a larger basis set $6311$++g(d,p)
was employed. To perform TDDFT calculations, the Gaussian16 software
package implements a framework based on an adiabatic approximation
in the frequency domain\citep{1996CPL...256..454B,10.1063/1.1508368,casida1998molecular,strat,VANCAILLIE1999249,VANCAILLIE2000159,scalmani}.
We employed an ``ultrafine'' numerical integration grid composed
of 99 radial layers and 590 angular divisions per layer, along with
an energy convergence limit set to 10\textsuperscript{-}\textsuperscript{6}
eV. No structural relaxation was conducted for the excited states,
instead, we computed 30 vertical excitation energies for all TQDs.
As far as the choice of exchange-correlational functional for the
TDDFT calculations is concerned, for the smallest structure, i.e.,
TQD-16 which was also used to benchmark the PPP parameters, we used
three functionals namely, B3LYP, PBE\citep{PhysRevLett.100.136406},
and HSE06\citep{heyd2003hybrid}. Because the agreement among the
spectra computed using the three functionals for TQD-16 was excellent,
for larger TQDs we only used the B3LYP functional.

\subsection{PPP model based approach}

\label{subsec:PPP-model}

Based on our TDDFT results, we have parameterized an effective $\pi$-
electron model Hamiltonian called Pariser-Parr-Pople (PPP) model\citep{Pople1953,pariser1953-ppp}
for TQDs considered in this work. The PPP model was originally proposed
for aromatic hydrocarbons containing benzonoid rings with $\pi$-conjugated
electrons, and later on was extended to other $\pi$-electron systems
such as linear polyenes, polyacetylene,\citep{baeriswyl1992conjugated},
fullerenes, graphene quantum dots and nanoribbons.\citep{Gundra2013}
The underlying assumption of the PPP model is the so-called $\sigma-\pi$
separation of electrons, in which the $\sigma$ electrons are assumed
to provide the structural backbone of the system but do not contribute
to the low-lying excited states. On the other hand, the excitations
involving the itinerant $\pi$ electrons close to the Fermi level,
determine the optical response of the system. The novelty of our work
lies in the fact that to the best of our knowledge, PPP model has
so far not been used for $\pi$-electron systems consisting of non-benzenoid
rings such as tetragons and octagons that form TQDs. In the second-quantized
notation, the PPP model can be written as

\begin{equation}
H=-\sum_{i,j,\sigma}t_{ij}(c_{i\sigma}^{\dagger}c_{j\sigma}+c_{j\sigma}^{\dagger}c_{i\sigma})+U\sum_{i}n_{i\uparrow}n_{i\downarrow}+\sum_{i<j}V_{ij}(n_{i}-1)(n_{j}-1),\label{eq:1}
\end{equation}

where creation (annihilation) operator $c_{i\sigma}^{\dagger}$($c_{i\sigma}$)
creates (annihilates) a $\pi$ - electron with spin $\sigma$ at the
$i^{th}$ carbon atom (site), while $n_{i}=\sum_{\sigma}c_{i\sigma}^{\dagger}c_{i\sigma}$
represents the number density of $\pi$ - electrons on the $i$-th
site. In Eq. \ref{eq:1}, the repulsive interaction between two electrons
is described by the second and third terms, where $U$ and $V_{ij}$,
respectively, denote the strengths of the onsite and the long-range
Coulomb interactions.

In our calculations, electron hopping is restricted to the nearest
neighboring sites, $i$ and $j$ (say), and is computed using the
linear interpolation formula $t_{ij}=t_{0}+3.2(R_{ij}-1.397),$with
$t_{0}=-2.4$ eV, and $R_{ij}$ the distance between the two sites
in $\text{\AA}$ units\citep{soos-polyenes-1982}. To reduce the number
of parameters, the Coulomb interaction $V_{ij}$ between the electrons
located on sites $i$ and $j$, is parameterized based on the Ohno
relationship \citep{Ohno1964} with a slight modification by Chandross
and Mazumdar\citep{Chandross1997} to account for the screening effects

\begin{equation}
V_{ij}=\frac{U}{\kappa_{ij}(1+0.6117R_{ij}^{2})^{\frac{1}{2}}},\label{eq:2}
\end{equation}

where $\kappa_{ij}$ represents the dielectric constant of the system
which takes into account the screening effects, while the rest of
the terms have already been discussed above. We have performed calculations
using two sets of Coulomb parameters: (a) the ``screened parameters''
proposed by Chandross and Mazumdar\citep{Chandross1997}with $U=8.0$
eV, $\kappa_{i,j}=2.0(i\ne j)$ and $\kappa_{i,i}=1.0$, and (b) the
``standard parameters'' with $U=11.13$ eV and $\kappa_{i,j}=1.0$,
proposed originally by Ohno\citep{Ohno1964}.{} We observe that our
calculations employing the screened parameters are in good agreement
with the first-principles TDDFT results.

We start our calculations by computing the self-consistent solutions
at the Restricted Hartree-Fock (RHF) level within the PPP model, using
a code developed in our group \citep{Sony2010}. We used the convergence
thresholds of $10^{-8}$ eV on the total Hartree-Fock energy, and
$10^{-5}$ eV on the single-particle energies. The molecular orbitals
(MOs) thus obtained are used to represent the Hartree-Fock ground
state with the electrons occupying the lowest-energy orbitals consistent
with the aufbau principle. The excited states of the system can also
be represented using these orbitals by promoting electrons from the
occupied to the virtual states. In the second step, we transform the
PPP Hamiltonian from the site representation to the MO representation
and incorporate the electron-correlation effects by performing the
CI calculations. While the single-reference approaches such as the
singles-doubles CI (SDCI) and quadruple CI (QCI) account for the correlation
effects in the ground state quite well, however, for the excited states
they may not perform that well. Therefore, an iterative multi-reference
singles-doubles CI (MRSDCI)\citep{buenker1978applicability,buenker1974theor}
targeting a chosen set of excited states is employed for the calculations
of excited states followed by the calculation of the optical absorption
spectra\citep{shukla2002correlated,PhysRevB.69.165218,chakraborty2013pariser,Basak2015,pritam-ppp}.

Thereupon, transition dipole matrix elements between the ground and
excited states are computed from the CI wave functions that are subsequently
used to compute the optical absorption spectrum, $\sigma(\omega)$,
of the TQD under consideration using the formula

\begin{equation}
\sigma(\omega)=4\pi\alpha\sum_{i}\frac{\omega_{i0}|\langle i|\hat{e}\cdot{\bf r}|0\rangle|^{2}\gamma}{(\omega_{i0}-\omega)^{2}+\gamma^{2}}.\label{eq:sigma-omega}
\end{equation}
Above, $\omega$ is the frequency of incident photon, $\hat{e}$ denotes
its polarization direction, ${\bf r}$ is the position operator, $\alpha$
is the fine structure constant, $|0\rangle$ and $|i\rangle$ denote
the CI wave functions of the ground and the $i$-th excited state,
respectively, $\omega_{i0}$ is the corresponding frequency difference,
while $\gamma$ is a uniform line width corresponding to the Lorentzian
line shape.

\section{Results and discussion}

\label{sec:Results-and-discussions}

\subsection{Optimized Geometries}

\begin{figure}[H]
\centering{}\includegraphics[scale=3]{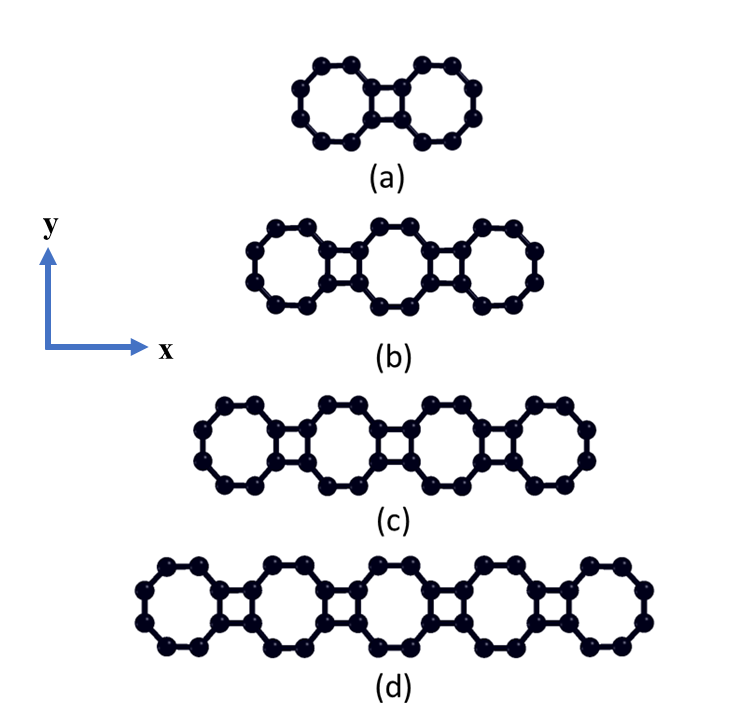}\caption{Schematic illustrations of T-graphene quantum dots investigated in
this study: (a) TQD-16, (b) TQD-24, (c) TQD-32, and (d) TQD-40. The
$x$-axis is taken to align with the conjugation direction, while
the $y$-axis is oriented perpendicular to it, within the plane of
the figure. The edge H atoms have not been shown.}
\label{fig:TQD_schematics} 
\end{figure}

The schematic structures of TQDs considered in this work are shown
in fig. \ref{fig:TQD_schematics}, with each quantum dot as TQD-$n$,
where $n$ is the total number of carbon atoms in the system. All
the TQDs are considered to lie in the $x-y$ plane, with the long
axis of the TQD assumed to be along the $x$- axis and the short one
along the $y$ - axis. The ideal bond angle involving the carbon atoms
in an octa-ring is $135^{o}$, while that in a tetra-ring is $90^{o}$.
Our optimized bond angles are found to be close to these ideal values,
while the carbon-carbon bond lengths of TQDs vary around the ideal
value of the C-C bond length of $1.4$ $\text{\AA}$. For the case
of TQD-16, the bond lengths range from a minimum value of $\text{1.378}$
$\text{\AA}$ to the maximum value $1.485\text{ \AA}$, with the bond
angles within the octagon varying from $130.11^{o}$to $137.92^{o}$,
while those within the tetraring remain close to $90^{o}$. For TQD-24,\textcolor{red}{{}
}the bond lengths span from a minimum of $1.39\text{\AA}$ to a maximum
of $1.51$$\text{\AA}$, while the range of optimized bond angles
inside the octagons is $130.20^{o}$ to $139.64^{o}$, and in the
squares it is $88.87^{o}$ and $91.14^{o}$. Similarly for TQD-32,
the C-C bond lengths range from a minimum $1.36$ $\text{\AA}$ to
a maximum $1.50\text{\AA}$, and the bond angles inside octa-rings
lie in the range $127.77^{o}$--$139.95^{o},$while in the tetra-rings
they range from $88.73^{o}$ to $91.28^{o}$. Lastly, for TQD-40,
the C-C bond lengths vary from 1.386 $\text{\AA}$ to 1.489 $\text{\AA}$,
with the bond angles in the octagons varying between $129.87^{o}$
and $139.89^{o}$, and in the tetra-rings between $88.83^{o}$ and
$91.17^{o}$. The optimized C-H bond lengths for all the TQDs were
in the narrow range 1.086--1.088 $\text{\AA}$. These optimized coordinates
of carbon atoms which are obtained from the first-principles DFT calculations
were used to perform the PPP model calculations calculations. Note
that in the PPP model, hydrogen atoms are ignored because of their
participation only in $\sigma$ bonds.\textcolor{magenta}{{} }

From the optimized values of the C-C bond lengths and the corresponding
bond angles, it is obvious that the octa- and tetra-rings of TQDs
are significantly different from regular octagons and squares. Nevertheless,
all the TQDs exhibit high symmetry corresponding to the $D_{2h}$
point group, with the ground state belonging to the $^{1}A_{g}$ irreducible
representation (irrep) and the dipole-allowed one-photon states belonging
to $^{1}B_{2u}$ ($y$-polarized), and $^{1}B_{3u}$($x$-polarized)
irreps. On further examining the symmetries of various MOs, we found
that both the HOMO and LUMO of these TQDs have the same inversion
symmetry i.e., $gerade$ type for TQD-16 and TQD-24, and $ungerade$
type for the remaining two TQDs, making the HOMO-LUMO optical transition
dipole forbidden in them. After the geometry optimization, vibrational-frequency
analysis was performed to verify the dynamic stability of TQDs, and
all the frequencies were found to be real. As far as the thermodynamic
stablility is concerned, T-graphene has 0.52 eV higher cohesive energy
per carbon atom than graphene\citep{Liu2012}. To check the thermodynamic
stability of TQDs, cohesive energies per atom of these quantum dots
were calculated as presented in Table \ref{tab:Cohesive-energy-per}.We
note that all the TQDs show negative cohesive energies, indicating
high thermodynamic stability of these structures, and, furthermore,
on increasing the number of octagons, cohesive energy of TQDs become
more negative, indicating that the infinite length TQD nanoribbons
are stable, in agreement with the previous studies\citep{C2CP41464C,Dai_2014}.
Next, we compare the calculated cohesive energies per atom of TQDs
to those of graphene quantum dots (GQDs) of various shapes and sizes
shown in Fig. S1 of the Supporting Information (SI). These GQDs are
well-known hydrocarbons which are stable and have been experimentally
synthesized. From the cohesive energies per atom of these GQDs calculated
at the same level of theory (B3LYP/6-311++G(p,d)) and presented in
Table S1 of SI, it is obvious that the two sets of cohesive energies
are quantitatively similar, indicating that TQDs can also be experimentally
synthesized.

\begin{table}[H]
\begin{centering}
\begin{tabular}{ccc}
\toprule 
System  & Molecular formula  & Cohesive energy per atom (eV)\tabularnewline
\midrule
\midrule 
TQD-16  & $C_{16}H_{12}$  & $-6.0558$\tabularnewline
TQD-24  & $C_{24}H_{16}$  & $-6.1965$\tabularnewline
TQD-32  & $C_{32}H_{20}$  & $-6.3029$\tabularnewline
TQD-40  & $C_{40}H_{24}$  & $-6.3660$\tabularnewline
\bottomrule
\end{tabular}
\par\end{centering}
\caption{Cohesive energy per atom of T-graphene quantum dots .}
\label{tab:Cohesive-energy-per} 
\end{table}

\subsection{Parameterization of the effective $\pi$- electron Hamiltoni an for
TQD-16}

In what follows, first we perform a series of calculations of the
low-lying excited states and optical absorption spectrum of the smallest
TQD, i.e, TQD-16, both using the first-principles TDDFT as well as
PPP-model-based CI (PPP-CI, henceforth) approach using various Coulomb
parameters. By comparing the results of the two sets of calculations,
we will determine which Coulomb parameters in the effective $\pi$-electron
Hamiltonian PPP-model lead to the best agreement with the TDDFT results.
Once the Coulomb parameters for the PPP-model for the case of TQD-16
are fixed, we will use the same set of parameters to perform calculations
on the longer TQDs presented in the subsequent sections.

As mentioned in the previous section, both in our first-princples
DFT as well PPP model-based Hartree-Fock calculations on TQD-16, HOMO
and LUMO orbitals have the same inversion symmetry, making the HOMO$\rightarrow$LUMO
optical transition dipole forbidden. Nevertheless, a low-lying excited
state whose many-particle wave function is dominated by the configuration
corresponding to the HOMO-LUMO virtual transition $|H\rightarrow L\rangle$
will exist. As far as the symmetry of this state is concerned, in
TQD-16 and TQD-24, HOMO and LUMO belong to $A_{g}$ and $B_{1g}$
irreps, respectively, so that $|H\rightarrow L\rangle$ state will
be of $B_{1g}$ symmetry. Similarly for TQD-32 and TQD-40, HOMO and
LUMO have $B_{3u}$ and $B_{2u}$ symmetries, respectively, due to
which $|H\rightarrow L\rangle$ state again belongs to the $B_{1g}$
symmetry. We call the excitation energy of this state as the HOMO-LUMO
(H-L) gap or $E_{HL}$, and calculate it both using both the TDDFT
and the PPP-CI methods. The CI method employed for these calculations
is quadruple-CI (QCI) method which includes up to four-electron excitations
with respect to the given reference state, thus including the electron-correlation
quite accurately.

Similarly, we compute the optical absorption spectra of TQD-16 using
the TDDFT approach, and compare those to the ones calculated using
the PPP-CI method and various Coulomb parameters.

Our TDDFT calculations are performed employing a triple-valence-zeta
basis along with the diffuse and polarization functions, i.e., $6311++g(d,p)$
basis set, and three exchange-correlation functionals, namely B3LYP,
PBE\citep{PhysRevLett.100.136406}, and HSE06\citep{heyd2003hybrid}.
For PPP-CI calculations we used two sets of Coulomb parameters mentioned
in the previous section, i.e., the standard and screened.

\subsubsection{HOMO-LUMO gap}

\label{subsec:HOMO-LUMO-gap}

In Table \ref{tab:HOMO-LUMO-gap-of-1}, we present the values of $E_{HL}$
computed at different levels of theory, along with the coefficient
of the $|H\rightarrow L\rangle$ configuration in the corresponding
many-electron wave function $|\psi_{HL}\rangle$. We note that: (a)
in the PPP-CI calculations performed using the QCI approach, the coefficients
of the $|H\rightarrow L\rangle$ configuration are close to 0.8, while
those in the TDDFT wave functions are close to 0.7, and (b) the values
of $E_{HL}$ obtained using the PPP-CI approach are significantly
larger than those computed using the TDDFT method. However, the values
of $E_{HL}$ and the coefficients of $|H\rightarrow L\rangle$ configuration
obtained using the TDDFT approach employing three different functionals
(B3LYP/HSE06/PBE) are in very good agreement with each other. We also
note that in the TDDFT calculations, the nature of HOMO/LUMO is $\pi/\pi^{*}$,
therefore, the state under consideration belongs to the class of $\pi\rightarrow\pi^{*}$
excitation, even though it is not optically accessible from the ground
state. The wave functions and the excitation energies of this state
for longer TQDs are presented in Table S13 of the Supporting Information
(SI).

\begin{table}[H]
\centering{}%
\begin{tabular}{ccc}
\toprule 
\multirow{1}{*}{Method} & \multicolumn{2}{c}{$|\psi_{HL}\rangle$ State}\tabularnewline
\midrule 
 & $E_{HL}$ (eV)  & Coefficient of $|H\rightarrow L\rangle$ configuration\tabularnewline
PPP-CI (scr)  & $2.18$  & 0.826\tabularnewline
PPP-CI (std)  & $1.80$  & $0.811$\tabularnewline
TDDFT (B3LYP)  & $1.50$  & $0.709$\tabularnewline
TDDFT (HSE06)  & $1.53$  & $0.709$\tabularnewline
TDDFT (PBE)  & $1.49$  & $0.708$\tabularnewline
\bottomrule
\end{tabular}\caption{The values of $E_{HL}$ of TQD-16 computed using the PPP-CI method
as well as TDDFT method employing three different functionals, namely,
B3LYP, HSE06, and PBE. The PPP-CI calculations were performed using
the QCI approach, employing both the standard and screened Coulomb
parameters.}
\label{tab:HOMO-LUMO-gap-of-1} 
\end{table}

\subsubsection{TDDFT spectrum}

\begin{figure}[H]
\centering{}\includegraphics[scale=3]{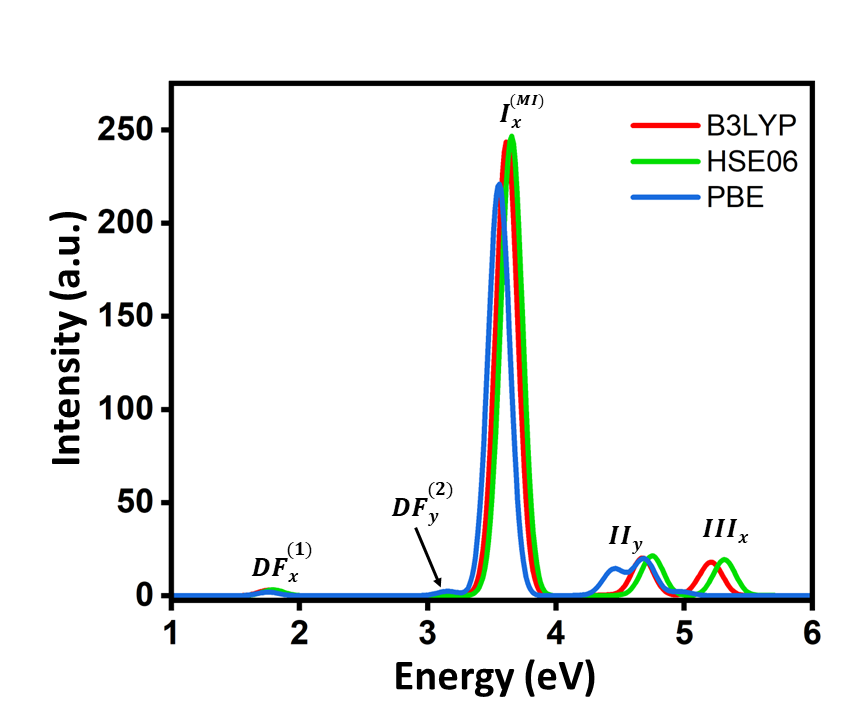}\caption{Comparison of optical absorption spectra of TQD-16 computed using
the first-principles TDDFT approach employing the B3LYP, PBE, and
HSE06 functionals, and the 6311++G(d,p) basis set. The subscript in
the peak labels indicate the polarization direction of the photon
involved.}
\label{fig:tqd-16-tddft-chi1} 
\end{figure}

The TDDFT optical absorption spectra of TQD-16, computed using 30
excited states and B3LYP, PBE, and HSE06 functionals, are presented
in Fig.\textcolor{red}{{} }\ref{fig:tqd-16-tddft-chi1}, while the
positions of the corresponding peaks and the dominant electronic transitions
are listed in Table \ref{tab:tqd16-tddft}. We note that the first
two peaks of the computed spectra are of symmetries $^{1}B_{3u}$
and$^{1}B_{2u}$ using all the three functionals, and have very low
intensities. As far as their peak locations are concerned, for the
first peak ($DF_{x}^{(1)}$), the results from all the three functionals
are in the range 1.76--1.79 eV, i.e., in very good agreement with
each other. However, for the second peak $(DF_{y}^{(2)})$ the value
computed using the PBE functional 3.16 eV is significantly smaller
than those computed using B3LYP (3.49 eV) and HSE06 (3.52 eV), which
are in a very good agreement with each other. Because the particle-hole
symmetry is absent in the TDDFT formalism, therefore, these peaks
are very weakly dipole-allowed unlike in the PPP-CI spectrum to be
discussed later, in which the corresponding peaks are completely dipole-forbidden
(DF) because of the particle-hole symmetry. Nevertheless, for easy
comparison of the TDDFT and PPP-CI spectra, we label these two peaks
as DF even for the TDDFT calculations. This peak (DF$_{y}^{(2)})$
is followed by the long-axis polarized ($^{1}B_{3u}$ symmetry) most
intense (MI) transition of the optical absorption spectra obtained
from TDDFT, with the peak locations 3.62 eV, 3.65 eV and 3.56 eV computed
using the B3LYP, HSE06, and PBE functionals, respectively. We note
that the peak location $(I_{x}^{(MI)})$\textcolor{magenta}{{} }computed
using the PBE functional is slightly red-shifted compared to the other
two which are within 0.03 eV of each other. After the MI peak, there
are two prominant peaks of $^{1}B_{2u}$ and $^{1}B_{3u}$ symmetries
(II$_{y}$ and III$_{x}$), subscripts indicate polarization direction).
The location of peak\textcolor{red}{{} } II$_{y}$ computed using
the B3LYP and HSE06 functionals is 4.68 eV and 4.75 eV, respectively.
While the location of this peak computed using the PBE functional
has a value 4.45 eV, which is shifted towards the lower energy compared
to B3LYP and HSE06 values. This is followed by peak\textcolor{red}{{}
}III$_{x}$, whose location computed from B3LYP, HSE06 and PBE functionals
is around 5.22 eV, 5.31 eV and 4.68 eV, respectively. Here it is noteworthy
that all the peaks of the absorption spectra, except the first peak,
computed from PBE functional are redshifted compared to the spectra
computed from B3LYP and HSE06 functionals.

We have further verified that the orbitals involved in the dominant
electronic excitations contributing to various peaks (see Table \ref{tab:tqd16-tddft})
are of $\pi/\pi^{*}$ type, therefore, it is fully justified to construct
an effective $\pi$-electron Hamiltonian for TQD-16 aimed at describing
its optical properties. We will see that this result holds for longer
TQDs as well. We further note that the spectra computed using the
B3LYP and HSE06 functionals are in very good agreement with each other
therefore on the longer TQDs we have performed TDDFT calculations
using only the B3LYP functional.

\begin{table}[H]
\begin{centering}
\begin{tabular}{cccccc}
\toprule 
 & \multicolumn{3}{c}{\begin{cellvarwidth}[t]
\centering
 {[}t{]} TDDFT Excitation Energies

(eV) 
\end{cellvarwidth}} &  & \tabularnewline
\midrule 
Symmetry  & B3LYP  & PBE  & HSE06  & \begin{cellvarwidth}[t]
\centering
 {[}t{]} Dominant Configurations

(B3LYP) 
\end{cellvarwidth} & Coefficient\tabularnewline
\midrule 
$^{1}B_{3u}$ ($DF_{x}^{(1)}$)  & $1.76$  & $1.76$  & $1.79$  & $|H-1\rightarrow L\rangle$  & 0.5604\tabularnewline
 &  &  &  & $|H\rightarrow L+1\rangle$  & -0.4343\tabularnewline
$^{1}B_{2u}$ ($DF_{y}^{(2)}$)  & $3.49$  & $3.16$  & $3.52$  & $|H-2\rightarrow L\rangle$  & 0.6603\tabularnewline
 &  &  &  & $|H\rightarrow L+2\rangle$  & -0.2450\tabularnewline
$^{1}B_{3u}$ ($I_{x}^{(MI)}$)  & $3.62$  & $3.56$  & $3.65$  & $|H\rightarrow L+1\rangle$  & 0.5728\tabularnewline
 &  &  &  & $|H-1\rightarrow L\rangle$  & 0.4600\tabularnewline
$^{1}B_{2u}$ $(II_{y})$  & $4.68$  & $4.45$  & $4.75$  & $|H\rightarrow L+2\rangle$  & 0.6120\tabularnewline
 &  &  &  & $|H-1\rightarrow L+8\rangle$  & 0.2589\tabularnewline
 &  &  &  & $|H-2\rightarrow L\rangle$  & 0.2042\tabularnewline
 &  &  &  & $|H-3\rightarrow L+1\rangle$  & 0.1211\tabularnewline
$^{1}B_{3u}$$(III_{x})$  & \multirow{1}{*}{$5.22$} & \multirow{1}{*}{$4.68$} & \multirow{1}{*}{$5.31$} & $|H-4\rightarrow L\rangle$  & 0.6834\tabularnewline
\bottomrule
\end{tabular}
\par\end{centering}
\caption{Comparison of the positions of the peaks (in eV) of the linear optical
absorption spectra of TQD-16 computed using the first-principles TDDFT
method and B3LYP, PBE, and HSE06 functionals. The weak peaks before
the maximum-intensity (MI) peak are labeled as ``dipole forbidden''
(DF) because their counterparts in the PPP-CI model are dipole forbidden
due to the electron-hole symmetry. The subscript in the peak labels
indicate the polarization direction of the photon involved.}
\label{tab:tqd16-tddft} 
\end{table}

\subsubsection{Tight-binding model spectrum}

\begin{figure}[H]
\centering{}\includegraphics[scale=3]{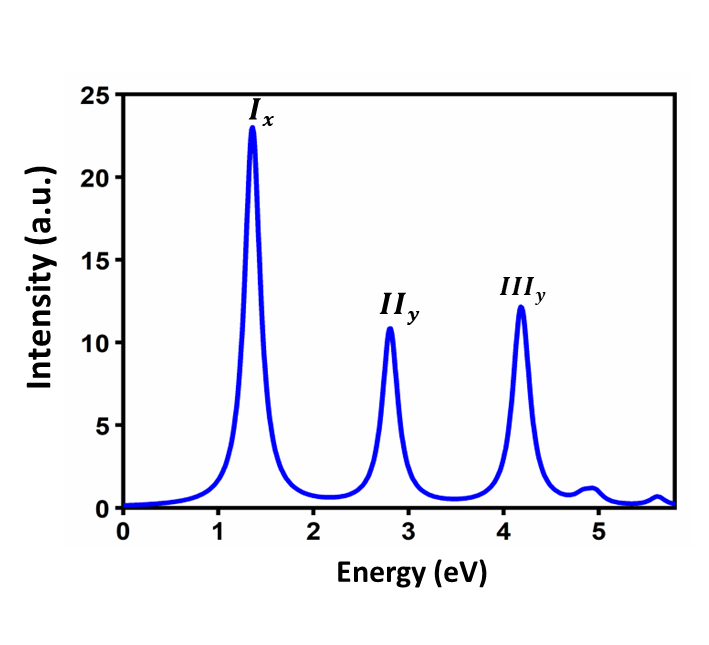}\caption{Optical absorption spectrum of TQD-16 computed using the tight-binding
model. The subscripts attached to the peak labels denote the directions
of polarization of the photons involved. The absorption spectra were
plotted assuming a uniform linewidth of 0.1 eV.}
\label{fig:TB-16} 
\end{figure}

Given the fact that the PPP model is a combination of two terms namely
one-body tight-binding (TB), and the two-body electron repulsion,
we first compute the absorption spectrum of TQD-16 at the TB level.
By comparing these results with the PPP-CI results which include the
contribution of both the terms, we will be able to understand the
influence of electron-correlation effects on the optical properties
of TQD-16. The optical absorption of TQD-16 computed using the TB
approach (see Fig. \ref{fig:TB-16}) contains three prominent peaks
followed by a couple of weak peaks at higher energies. Peak I$_{x}$
located at 1.35 eV is $x$-polarized corresponding to the transition
to an excited state of $B_{3u}$ symmetry, and involves orbital excitations
$|H\rightarrow L+1\rangle$ and $|H-1\rightarrow L\rangle$. Peak\textcolor{magenta}{{}
}II$_{y}$ located at 2.81 eV represents an optical transition to
a $B_{2u}$ symmetry state through orbital excitations $|H\rightarrow L+2\rangle$
and $|H-2\rightarrow L\rangle$, and a $y$-polarized photon. Finally,
peak III$_{y}$ at 4.17 eV again corresponds to a transition to a
$B_{2u}$ symmetry state through a $y$-polarized photon involving
$|H-1\rightarrow L+3\rangle$ and\textcolor{magenta}{{} }$|H-3\rightarrow L+1\rangle$
excitations. In the orbital transitions we note electron-hole (e-h)
symmetry which is a consequence of the fact that TQDs are a bipartite
system being described by a nearest-neighbor TB model.

\subsubsection{PPP-CI spectrum }

\begin{figure}[H]
\centering{}\includegraphics[scale=3]{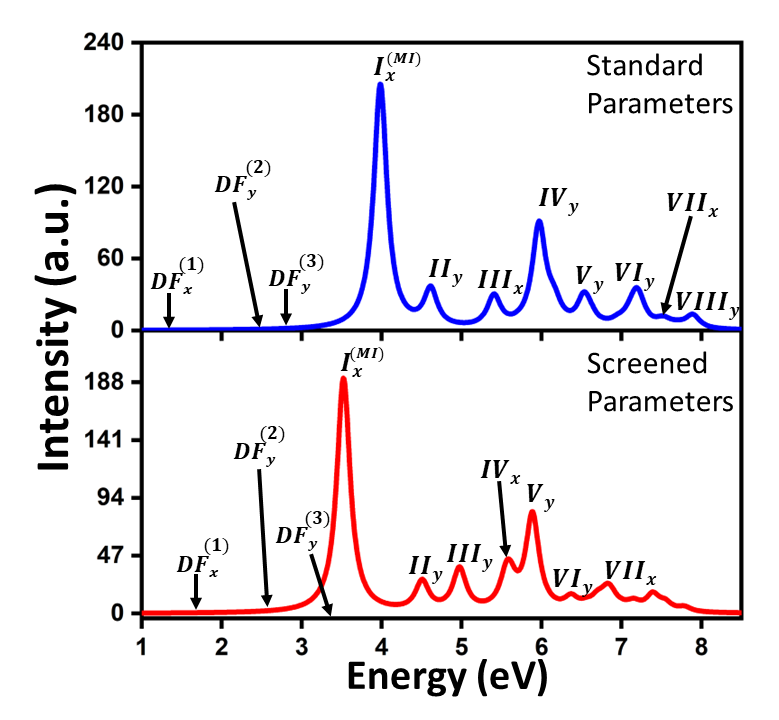}\caption{Optical absorption spectrum of TQD-16 computed employing the PPP-CI
approach, and the standard (top) and screened (bottom) Coulomb parameters.
The $1^{1}A_{g}$ ground state was computed using the QCI method,
while MRSDCI approach was utilized to compute all the excited states.\textcolor{magenta}{{}
}The absorption spectra were plotted assuming a uniform linewidth
of 0.1 eV.}
\label{fig:16-spec_1} 
\end{figure}

In Fig \ref{fig:16-spec_1}, we present the optical absorption spectra
of TQD-16 computed using the PPP-CI approach, employing the screened
and standard coulomb parameters, while in Table \ref{tab:Positions-of-the_tqd16}
we list the locations of those excited states and the dominant electronic
configurations contributing to their wave functions. For calculating
the spectra, the $1^{1}A_{g}$ ground state of the molecule was computed
using the QCI method, while the excited states were obtained using
our iterative MRSDCI method described in section \ref{subsec:PPP-model}.
Information containing more details about various peaks in the absorption
spectra is presented in tables S2 and S3 of SI. 

\begin{table}
\begin{tabular}{ccccc}
\toprule 
\multirow{1}{*}{Symmetry} & \multicolumn{4}{c}{PPP-CI}\tabularnewline
\midrule 
 & std (eV)  & scr (eV)  & Dominant configuration (scr)  & Coefficient (scr)\tabularnewline
\midrule
\midrule 
$B_{3u}$($DF_{x}^{(1)}$)  & $1.38$  & $1.66$  & $|H\rightarrow L+1\rangle-c.c$  & 0.5775\tabularnewline
\midrule 
$B_{2u}(DF_{y}^{(2)})$  & 2.58  & 2.72  & $|H\rightarrow L;H-1\rightarrow L\rangle-c.c$  & 0.4506\tabularnewline
 &  &  & $|H\rightarrow L+2\rangle-c.c$  & 0.3491\tabularnewline
\midrule 
$B_{2u}(DF_{y}^{(3)})$  & $2.79$  & $3.38$  & $|H\rightarrow L;H\rightarrow L+1\rangle-c.c$  & 0.5436\tabularnewline
\midrule 
$B_{3u}$ ($I_{x}^{(MI)}$)  & $3.98$  & $3.52$  & $|H\rightarrow L+1\rangle+c.c$  & 0.5880\tabularnewline
\midrule 
$B_{2u}$$(II_{y})$  & 4.61  & 4.50  & $|H\rightarrow L+2\rangle+c.c$  & 0.3728\tabularnewline
 &  &  & $|H\rightarrow L;H\rightarrow L+1\rangle+c.c$  & 0.2203\tabularnewline
\midrule 
$B_{3u}$$(III_{x}/IV_{x})$  & 5.36  & 5.58  & $|H-4\rightarrow L\rangle+c.c.$  & 0.4163\tabularnewline
 &  &  & $|H-2\rightarrow L+1;H-1\rightarrow L\rangle$  & 0.2630\tabularnewline
\midrule 
$B_{2u}(IV_{y}/V_{y})$  & 5.97  & 5.88  & $|H-1\rightarrow L+3\rangle-c.c$  & 0.5580\tabularnewline
\bottomrule
\end{tabular}\caption{Excitation energies of various peaks (in eV) in the linear optical
absorption spectra of TQD-16 obtained from the PPP-CI method using
the screened (scr) and standard (std) Coulomb parameters. MI denotes
the maximum-intensity peak and DF denotes the ``dipole forbidden''
peak. The subscript in the peak labels indicate the polarization direction
of the absorbed photon. In addition, the dominant configurations,
and their coefficients, in the screened-parameter CI expansions of
the excited states contributing to the peaks are also given.}
\label{tab:Positions-of-the_tqd16} 
\end{table}

From the PPP-CI spectrum, we find that the first excited state of
the system, which is of $B_{3u}$ symmetry (denoted as $DF_{x}^{(1)}$),
is dipole-forbidden because of the e-h symmetry, and is predicted
to be at 1.66 (1.38) eV by screened (standard) parameter calculations.
We note that the screened parameter value is in good agreement with
those computed using the TDDFT employing B3LYP (1.76 eV) and PBE ($1.76$
eV) functionals (see Table \ref{tab:tqd16-tddft}). We further note
that the good agreement between the PPP-CI (scr) results and TDDFT
ones also holds for the dominant orbital excitations involved in the
many-body wave functions of these excited states predicted to be $|H\rightarrow L+1\rangle$
and $|H-1\rightarrow L\rangle$ by both sets of calculations. It is
just that the coefficients of the configurations are equal and opposite
for the PPP-CI case because of the e-h symmetry, while they are unequal,
but still of opposite signs, for the TDDFT (B3LYP) calculations.

The next excited state $DF_{y}^{(2)}$ is of $B_{2u}$ symmetry which
is also strictly dipole forbidden in the PPP-CI calculations with
both sets of Coulomb parameters. Our screened (standard) parameter
calculations predict its location to be 2.78 (2.58) eV with the wave
function dominated by double excitations $|H\rightarrow L;H-1\rightarrow L\rangle-c.c$
and single excitations $|H\rightarrow L+2\rangle-c.c$. As far as
excitation energy is concerned, there is no counterpart of this state
in the TDDFT spectra, however, the wave function of state\textcolor{magenta}{{}
}$DF_{y}^{(2)}$\textcolor{magenta}{{} }of TDDFT (see Table \ref{tab:tqd16-tddft})
is also dominated by single excitations $|H\rightarrow L+2\rangle$
and $|H-2\rightarrow L\rangle$. Because the MRSDCI approach allows
for double excitations in its wave function expansion unlike the TDDFT
approach, therefore, because of them the energy of PPP-CI $DF_{y}^{(2)}$
state has shifted below that of the TDDFT $DF_{y}^{(2)}$\textcolor{magenta}{{}
}state.

This is followed by another state of $B_{2u}$ symmetry labeled $DF_{y}^{(3)}$
with a very weak transition dipole moment of 0.02 �(0.03 �) in
the PPP-CI screened (standard) parameter calculations. Because of
its almost negligible transition dipole moment, we have labeled it
as a dipole-forbidden state. Its excitation energy computed using
the screened (standard) parameters is 3.38 (2.79) while the configurations
contributing dominantly to the wave function are double excitations
such as $|H\rightarrow L;H\rightarrow L+1\rangle-c.c$, and several
others. The state with the closest excitation energy to this one in
the TDDFT spectrum is labeled\textcolor{magenta}{{} }$DF_{y}^{(2)}$
(see Table \ref{tab:tqd16-tddft}) and predicted to be at 3.49/3.16/3.52
eV using B3LYP/PBE/HSE06 functionals. We note that the screened parameter
excitation energy of 3.38 eV of the $DF_{y}^{(3)}$ is reasonably
close to the TDDFT values, however, wave functions predicted by PPP-CI
calculation dominated by the double excitations is completely different
from the one predicted by the TDDFT calculations which consists mainly
of the single excitations $|H\rightarrow L+2\rangle$ and $|H-2\rightarrow L\rangle$.
Therefore, in spite of good agreement between the PPP-CI (scr) and
TDDFT excitation energies, the nature of the corresponding states
when it comes to their wave functions is quite different.

The first peak of the absorption spectrum with a significant intensity
which is also its most intense (MI) peak $I_{x}^{(MI)}$ is predicted
at 3.52 (3.98) eV by the screened (standard) parameter calculations.
The peak is long-axis polarized corresponding to an excited state
of the $B_{3u}$ symmetry with the many-electron wave function mainly
composed of the $|H\rightarrow L+1\rangle+c.c$. single excitations.
On comparing it with the TDDFT results (see Table \ref{tab:tqd16-tddft}),
we find that the B3LYP excitation energy is 0.1 eV larger than our
screened parameter value, while the PBE value is just 0.04 eV higher.
Furthermore, the B3LYP wave function is also composed of $|H\rightarrow L+1\rangle$
and $|H-1\rightarrow L+1\rangle$ with almost equal coefficients of
the same sign. This aspect of the PPP-CI and TDDFT spectra is in excellent
agreement with the tight-binding absorption spectrum (see Fig. \ref{fig:TB-16})
which also predicts the most intense peak to be the first one caused
by the transitions $|H\rightarrow L+1\rangle$ and $|H-1\rightarrow L+1\rangle$.
However, the excitation energies predicted by the TB theory are much
lower as compared to those of PPP-CI and TDDFT because the TB theory
does not include the electron-correlation effects.

The next peak ($II_{y}$) in the PPP-CI spectrum is short-axis polarized
and due to a state of the $B_{2u}$ symmetry, just as in the TDDFT
spectrum (see Fig. \ref{fig:tqd-16-tddft-chi1}) Similar to the TDDFT
case, peak $II_{y}$ is also much weaker compared to the MI peak.
Its location computed the using screened(standard) parameters is 4.50
(4.61) eV which is in good agreement with TDDFT values using the PBE
(4.45 eV) and B3LYP (4.68 eV). We note that the screened parameter
value is about 0.2 eV lower than the B3LYP value and quite close to
the PBE value. The dominant single excitations contributing to the
CI wave function are $|H\rightarrow L+2\rangle+c.c$, which is in
perfect agreement with states contributing to the second peaks of
both the TDDFT and TB spectra. However, due to electron-correlation
effects, in the PPP-CI wave function, doubly-excited configurations
also make important contributions.

The final peak in the PPP-CI spectra that can be compared to one in
the TDDFT spectra is $IV_{x}$ ($III_{x}$) in the screened (standard)
parameter calculations corresponding to $B_{3u}$ symmetry located
at 5.58 (5.36) eV. As a matter of fact, in the screened parameter
calculations, two closely-spaced states located at 5.56 and 5.60 eV
contribute to this peak (see Table S2). The PPP-CI excitation energies
compare well with the predicted TDDFT location of $III_{x}$ peak
(see Fig. \ref{fig:tqd-16-tddft-chi1}) at 5.22 (5.31) eV using the
B3LYP (HSE06) functional. We note that TDDFT wave function of this
state is dominated by the single excitation $|H-4\rightarrow L\rangle$
(see Table \ref{tab:tqd16-tddft}) while the PPP-CI wave functions,
in addition to the single excitations $|H-4\rightarrow L\rangle+c.c.$,
also derives important contributions from the double excitations.

In addition to the peaks that can be correlated with the TDDFT peaks,
the PPP-CI spectrum contains several other peaks that are absent in
the TDDFT spectra. For example, there is no TDDFT counterpart of the
second most intense peak in the PPP-CI spectra which is due to a state
of $B_{2u}$ symmetry labeled $IV_{y}$ (standard) and $V_{y}$(screened)
located at 5.97 and 5.88 eV, respectively. The wave functions for
both the standard and the screened parameter calculations for these
states are dominated by the single excitations $|H-1\rightarrow L+3\rangle\pm c.c$,
along with contributions from doubly excited configurations (see Tables
S2 and S3). We note that the third peak of the TB spectrum (see Fig.
\ref{fig:TB-16}) is also $y$-polarized due to single excitations
|$H-1\rightarrow L+3\rangle$ and |$H-3\rightarrow L+1\rangle$. Therefore,
we conclude that this peak of the PPP-CI spectrum owes its origins
to the third peak obtained from the TB model, except that the PPP-CI
wave functions also have contributions due to doubly-excited configurations
due to the inclusion of the electron-correlation effects.

\begin{figure}[H]
\centering{}\includegraphics[scale=3]{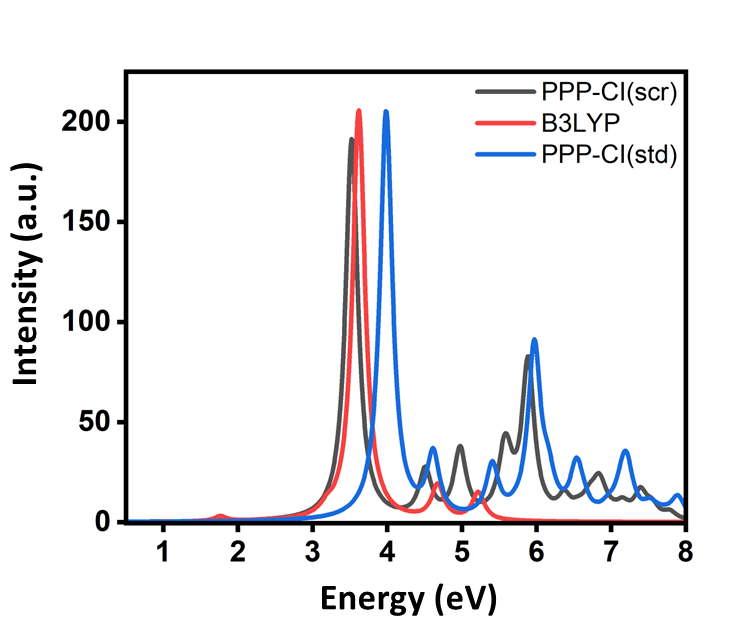}\caption{Comparison of the optical absorption spectra of TQD-16 calculated
using the PPP-CI method, and the standard (std) and screened (scr)
Coulomb parameters, with that computed using the first-principles
TDDFT approach, and B3LYP/$6311++G(d,p)$ level of theory. }
\label{fig:Comparison-of-optical} 
\end{figure}

Now that we have analyzed the nature of the low-lying excited states
of TQD-16 computed using TDDFT and PPP-CI methods in detail, the question
arises as to which set of results are in good agreement with each
other. Based on our discussion so far, it is clear that the PPP-CI
(screened) results for the optically allowed states are in very good
agreement with the TDDFT (B3LYP) results. This is further confirmed
by Fig. \ref{fig:Comparison-of-optical} in which we have plotted
the spectra computed using the PPP-CI method employing both the standard
and screened Coulomb parameters with the one computed using the TDDFT
(B3LYP) method. The close agreement between the PPP-CI (screened)
results and the TDDFT results is quite obvious. Therefore, we conclude
that screened parameters that were obtained originally for benzoid
hydrocarbons containing hexagons work well even for TQDs that contain
tetragons and octagons.

\subsection{Optical absorption in longer TQDs}

\begin{figure}[H]
\centering{}\includegraphics[scale=3]{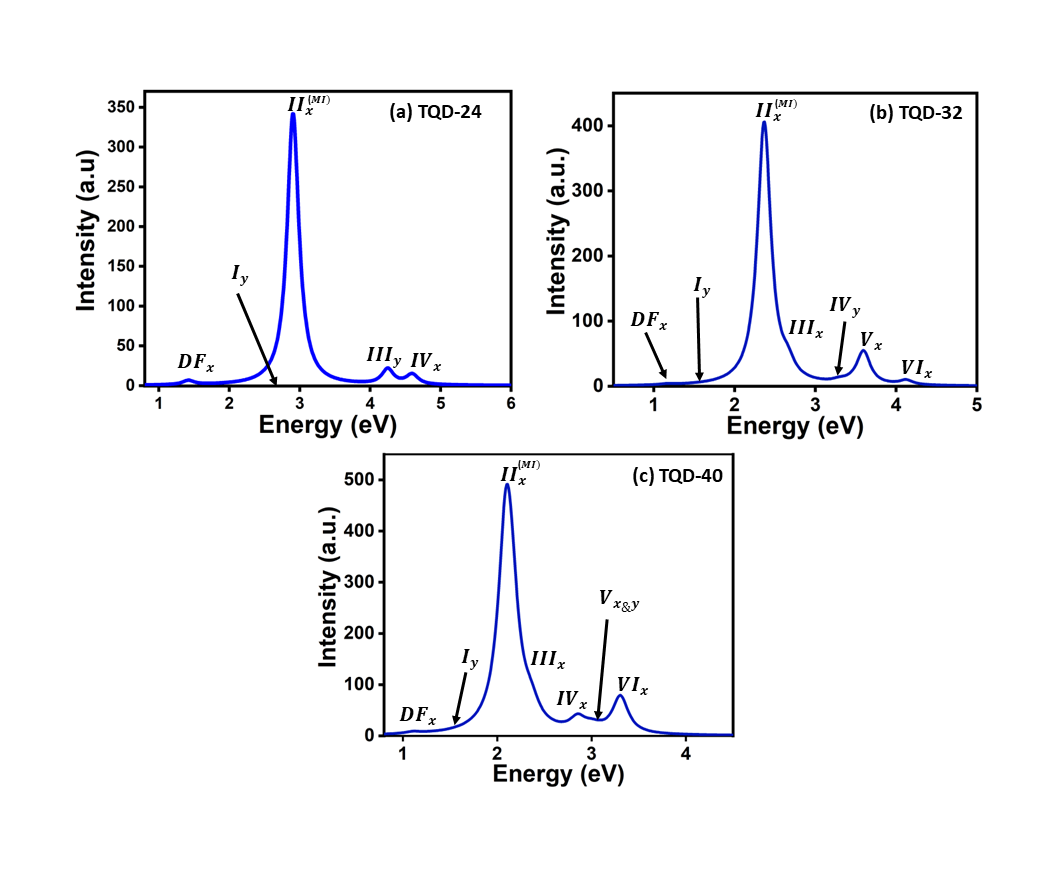}\caption{The first-principles TDDFT spectra of (a) TQD-24, (b) TQD-32, and
(c) TQD-40, calculated using the B3LYP functional, and $6311++G(d,p)$
basis set.}
\label{fig:tddft-for-longer-tqds} 
\end{figure}

\begin{figure}[H]
\centering{}\includegraphics[scale=3]{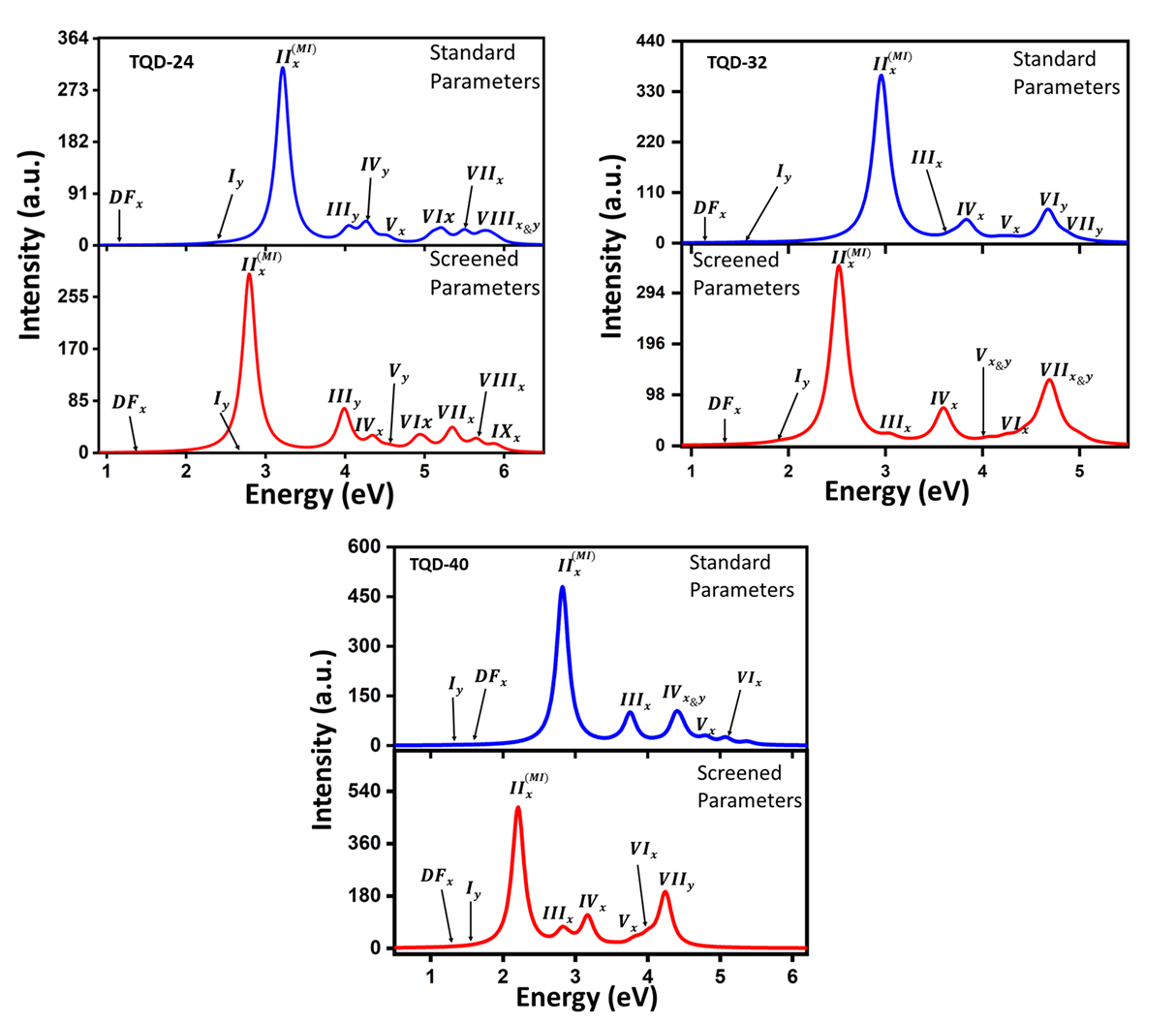}\caption{The PPP-CI spectra of TQD-24, TQD-32, and TQD-40, calculated using
the standard and screened Coulomb parameters. The absorption spectra
were plotted assuming a uniform linewidth of 0.1 eV.}
\label{fig:ppp-ci-for-longer-tqds} 
\end{figure}

In this section we discuss the PPP-CI and TDDFT results for the longer
TQDs, namely, TQD-24, TQD-32 and TQD-40, and analyze the general characteristics
of their optical absorption spectra. In Figs. \ref{fig:tddft-for-longer-tqds}
and \ref{fig:ppp-ci-for-longer-tqds} we present the spectra of the
longer TQDs computed using the TDDFT and PPP-CI approaches, respectively,
while Table \ref{tab:comparison_excited_state} contains the locations
of their most important peaks. Additionally, detailed information
about the excited states contributing to the spectra for the PPP-CI
calculations are presented in Tables S4--S9, and in Tables S10--S12
for the TDDFT calculations. Moreover, in Fig. S2 we compare the plots
of the spectra computed using the PPP-CI and TDDFT methods, for each
TQD separately.

Based on this, we note the following general trends: 
\begin{enumerate}
\item For all types of calculations, the spectra are redshifted with the
increasing size of the TQDs. 
\item For a given TQD, the screened parameter based PPP-CI spectra are redshifted
compared to the standard parameter results (see Fig. \ref{fig:ppp-ci-for-longer-tqds}) 
\item For a given TQD, overall agreement TDDFT calculations and the screened
parameter PPP-CI spectra is much superior as compared to the standard
parameter results (see Fig S2 of SI) 
\item Because of the e-h symmetry of the half-filled PPP model, in the PPP-CI
calculations a configuration, (say $|H-n\rightarrow L+m\rangle$,
where $m\mbox{ and }n$ are integers) and its e-h reversed counterpart
(called $c.c.$ by us, and for this case $|H-m\rightarrow L+n\rangle$)
contribute with coefficients of equal magnitude. However, that symmetry
is not observed in the TDDFT calculations for the longer TQDs. 
\item In our PPP model calculations, the HOMO and LUMO orbitals of the longer
TQDs have the same inversion symmetry, thereby making the $|H\rightarrow L\rangle$
transition dipole forbidden, in agreement with TQD-16. We note that
to be the case in the TDDFT calculations as well, except for the case
of TQD-24, for which $|H\rightarrow L\rangle$ transition is dipole
allowed. 
\item The first excited state of $B_{3u}$ symmetry ($DF_{x})$ is strictly
dipole-forbidden due to the e-h symmetry in the PPP-CI calculations,
while it is very weakly allowed in the TDDFT ones. The set of configurations
that contribute to this state are the same ones that contribute of
the MI transition $(II_{x}^{(MI)})$ is which is also of $B_{3u}$
symmetry, that is long-axis polarized. The orthogonality condition
forces the relative signs of the configurations to be opposite for
the $DF_{x}$ and $II_{x}^{(MI)}$ states leading to vanishing transition
dipole moment for the former, and extremely large one for the latter.
For the PPP-CI calculations the configurations are $|H\rightarrow L+1\rangle/|H-1\rightarrow L\rangle$
for all the TQDs, while in the TDDFT calculations they were $|H-1\rightarrow L+1\rangle/|H\rightarrow L\rangle$
for TQD-24, and $|H-1\rightarrow L\rangle/|H\rightarrow L+2\rangle$
for TQD-32 and TQD-40. 
\item The location of $DF_{x}$ computed from TDDFT spectrum is in good
agreement with that computed from PPP-CI method using screened parameters,
in all the longer TQDs except in TQD-32, where the agreement is better
between TDDFT (1.16 eV) and PPP-CI (1.19 eV) using the standard parameters.
The screened parameters PPP-CI calculations predict a value 0.18 eV
less than the TDDFT value. 
\item The first allowed transition computed from PPP-CI methodology is a
very weak peak ($I_{y}$ ) corresponding to the absorption of $y-$polarized
photon in all the longer TQDs. In the TDDFT spectra (see Fig. \ref{fig:tddft-for-longer-tqds}),
this peak is almost dipole forbidden because of its negligible oscillator
strength. The wave function of $I_{y}$ state in the PPP-CI calculations
is dominated by the single excitations $|H\rightarrow L+2\rangle\pm c.c$
for all the TQDs, except in TQD-24, where it is dominated by a doubly
excited configurations $|H\rightarrow L;H\rightarrow L+1\rangle\pm c.c.$.
In case of the TDDFT calculations (see Tables S10--S12), for TQD-24
the dominant configuration is $|H\rightarrow L+2\rangle$, while for
TQD-32 and TQD-40 it is $|H-2\rightarrow L\rangle$. 
\item The locations of $I_{y}$ predicted by TDDFT and PPP-CI (screened)
calculations are generally in quite good agreement with each other
except for the case of TQD-32 for which standard parameter results
are in almost exact agreement with the TDDFT location. 
\item As far as the location of the most-intense $II_{x}^{(MI)}$ peak is
concerned, the agreement between the TDDFT and PPP-CI (screened) calculations
is quite good, while the standard-parameter-based PPP-CI results are
significantly blue shifted. 
\item The good agreement between PPP-CI (screened) and the TDDFT results
continues for several higher energy states beyond the most-intense
peak (see Table \ref{tab:comparison_excited_state}). 
\end{enumerate}
Based on the aforesaid observations, we conclude that we have achieved
the goal of parameterization of the PPP model for describing the optical
absorption spectra of TQDs, in coordination with the TDDFT calculations.
Our results suggest that the screened Coulomb parameters provide overall
better agreement with the TDDFT/B3LYP absorption spectra as compared
to the standard parameters. Therefore, it will be tremendously useful
if these structures can be synthesized in laboratory and their optical
spectra measured to determine the final accuracy of our predictions.

\begin{table}[H]
\begin{centering}
\begin{tabular}{ccccc}
\toprule 
\multirow{1}{*}{No. of atoms in TQD} & \multirow{1}{*}{Symmetry} & \multicolumn{1}{c}{First-principles} & \multicolumn{2}{c}{PPP-CI}\tabularnewline
 &  & \multirow{2}{*}{TDDFT} & \multirow{2}{*}{scr} & \multirow{2}{*}{std}\tabularnewline
 &  &  &  & \tabularnewline
\midrule 
24  & $^{1}B_{3u}$($DF_{x}$)  & $1.42$  & $1.35$  & $1.16$\tabularnewline
 & $^{1}B_{2u}$$(I_{y})$  & \multirow{1}{*}{$2.63$} & \multirow{1}{*}{$2.70$} & \multirow{1}{*}{$2.35$}\tabularnewline
 & $^{1}B_{3u}$ $(II_{x}^{(MI)})$  & $2.90$  & $2.79$  & $3.22$\tabularnewline
 & $^{1}B_{2u}(III_{y})$  & $4.10$  & $4.00$  & 3.96\tabularnewline
 & $^{1}B_{2u}(IV_{x})$  & $4.25$  & 4.35  & $4.25$\tabularnewline
 &  &  &  & \tabularnewline
32  & $^{1}B_{3u}$($DF_{x}$)  & $1.16$  & $1.34$  & $1.19$\tabularnewline
 & $^{1}B_{2u}$$(I_{y})$  & \multirow{1}{*}{$1.58$} & \multirow{1}{*}{$1.92$} & \multirow{1}{*}{$1.59$}\tabularnewline
 & $^{1}B_{3u}$($II_{x}^{(MI)}$)  & $2.37$  & $2.52$  & $2.96$\tabularnewline
 & $^{1}B_{3u}(III_{x})$  & 2.66  & 3.05  & 3.69\tabularnewline
 & $^{1}B_{3u}(IV_{x}/V_{x})$  & 3.60  & 3.60  & 3.83\tabularnewline
 &  &  &  & \tabularnewline
40  & $^{1}B_{3u}$($DF_{x}$)  & $1.10$  & $1.25$  & $1.58$\tabularnewline
 & $^{1}B_{2u}$$(I_{y})$  & \multirow{1}{*}{$1.65$} & \multirow{1}{*}{$1.72$} & \multirow{1}{*}{$1.37$}\tabularnewline
 & $^{1}B_{3u}$$(II_{x}^{(MI)})$  & $2.09$  & $2.21$  & $2.82$\tabularnewline
 & $^{1}B_{3u}(III_{x})$  & 2.36  & $2.81$  & 3.75\tabularnewline
\bottomrule
\end{tabular}
\par\end{centering}
\caption{Comparison of position (in eV) of linear absorption peaks of TQDs
obtained from the First-Principles TDDFT using B3LYP with our PPP-CI
results. Dipole forbidden state and most intense peak are denoted
by DF and MI, respectively.}
\label{tab:comparison_excited_state} 
\end{table}

.

\subsection{Lowest Triplet State and the Singlet-Triplet Gap}

In order to understand low-lying spin excitations of TQDs, we have
calculated their spin gap, i.e., the excitation energy of their lowest
triplet state with respect to the ground state. The lowest triplet
state has the same dominant orbital excitation $|H\rightarrow L\rangle$
which determines the HOMO-LUMO gap (see section \ref{subsec:HOMO-LUMO-gap}),
except that the spin of the hole in the HOMO couples with that of
the the electron in the LUMO, to yield a spin triplet. Therefore,
it is computed as $\Delta E_{ST}=E(1^{3}B_{1g})-E(1^{1}A_{g})$ for
all the TQDs. In Table \ref{tab:Spin-gap} we present the values of
the spin gap for all the considered TQDs, calculated using the PPP-CI
approach. We note that irrespective of the Coulomb parameters used
in the calculations for all TQDs, the spin gap is smaller than 1 eV.
However, with respect to the size of the TQD, we see a more complicated
behavior: (a) the spin gap fluctuates as the length increases for
the screened parameter calculations, and (b) with the standard parameter
calculations initially the gap is constant close to 0.8 eV for TQD-16,
TQD-24, and TQD-32, however, for TQD-40 it increases by about 0.1
eV. The detailed wave functions of the $T_{1}$ state ($1^{3}B_{1g}$
state) for all the TQDs are presented in Table S14, and we note that
this state is dominated by the excitation $|H\rightarrow L\rangle$
similar to the singlet $|\psi_{HL}\rangle$ state (see Table S13).
However, the magnitude of the coefficient of the $|H\rightarrow L\rangle$
in $T_{1}$ is significantly smaller than that in the $|\psi_{HL}\rangle$
state, indicating that the $T_{1}$ state wave function has comparatively
more open-shell character driven by electron-correlation effects.
If we compare the energies of the spin gap with the HOMO-LUMO gap
corresponding to the $|\psi_{HL}\rangle$ state (see Table S13), and
find that $E_{HL}$ is several times larger than $\Delta E_{ST}$.
Because the dominant orbital excitations involved in both the states
are the same as mentioned above, the significant lowering of the triplet
state compared to the corresponding singlet state again testifies
to the influence of stronger electron-correlation effects in the triplet
manifold.

\begin{table}[H]
\begin{centering}
\begin{tabular}{ccc}
\toprule 
\multirow{1}{*}{System} & \multicolumn{2}{c}{Spin gap ($\Delta E_{ST}$)}\tabularnewline
\midrule 
 & Screened  & Standard\tabularnewline
\midrule
\midrule 
TQD-16  & $0.66$  & $0.76$\tabularnewline
TQD-24  & $0.75$  & $0.76$\tabularnewline
TQD-32  & $0.55$  & $0.78$\tabularnewline
TQD-40  & $0.82$  & $0.90$\tabularnewline
\bottomrule
\end{tabular}
\par\end{centering}
\centering{}\caption{The singlet-triplet energy gap (in eV), i.e., the spin gap, of TQDs
computed using the PPP-CI approach, and the standard and screened
Coulomb parameters.}
\label{tab:Spin-gap} 
\end{table}

If we compare the spin gap of each quantum dot with the value of the
optical gap corresponding to the MI peak, we find the relationship
$E_{MI}>n\Delta E_{ST}$ for all the TQDs, with $n$ ranging from
2--5. This suggests that, from an energetic point of view, in TQDs
the MI state has the potential to undergo singlet fission into two
or more triplets.\citep{smith-michl-singlet-fission} If that is the
case experimentally, TQDs may prove to be very useful in photovoltaics.

\subsection{Are TQDs photoluminescent?}

\begin{figure}
\begin{centering}
\includegraphics[width=8cm]{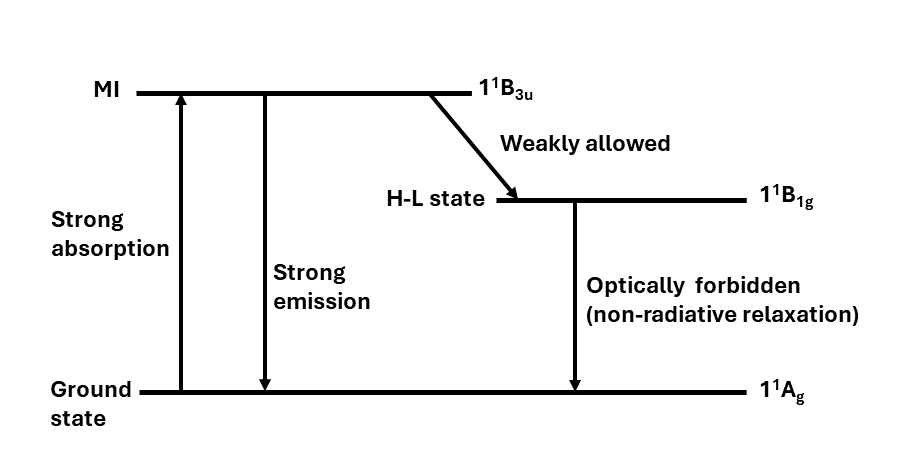} 
\par\end{centering}
\caption{This figure shows the low-lying spin-singlet excited states and their
transition dipole moments. Because of the strong dipole coupling of
the\textcolor{red}{{} }$1^{1}B_{3u}$\textcolor{magenta}{{} }state
to the ground state, and weak coupling to the $1^{1}B_{1g}$ ($|\psi_{HL}\rangle$)
state, an optically pumped TQD will decay predominantly to the ground
state, making it photoluminescent. On the other hand, the $1^{1}B_{1g}$
state will not have a large population, but will have a large lifetime
because only nonradiative decay to the ground state is allowed. }
\label{fig:PL} 
\end{figure}

The calculated optical absorption spectra of TQDs reveal that the
first intense absorption with a large oscillator strength is to the
state of $1^{1}B_{3u}$, the peak corresponding to which is labeled
as the MI peak. The question arises whether a TQD which has been optically
pumped to the MI state will decay back to the ground state or to the
$|\psi_{HL}\rangle$ state of$^{1}B_{1g}$ symmetry to which also
the $1^{1}B_{3u}\rightarrow1^{1}B_{1g}$ transition is dipole allowed.
Obviously, this will depend on the relative strength of the transition
dipole corresponding to the $1^{1}B_{3u}\rightarrow1^{1}B_{1g}$ transition
as compared to the decay to the ground state, i.e., $1^{1}B_{3u}\rightarrow1^{1}A_{g}$
transition. To determine this, we computed the transition dipole for
the $1^{1}B_{3u}\rightarrow1^{1}B_{1g}$ transition at the PPP-CI
level using both the screened and standard parameters and found it
to be exactly zero because of the e-h selection rules. As discussed
previously, because e-h symmetry is an approximate one, therefore,
in real life we expect this transition to be weakly allowed with a
very small oscillator strength (see Fig. \ref{fig:PL}). As a result,
the dominant decay from the $^{1}B_{3u}$ state will be to the ground
state, making the material photoluminescent, a property which may
prove to be useful in optoelectronics. The case in which the weak
decay to the $1^{1}B_{1g}$ state occurs, the TQD will be able to
reach the $1^{1}A_{g}$ only by nonradiative means because the transition
is dipole-forbidden because the two states have the same inversion
symmetry. Therefore, we expect this $|\psi_{HL}\rangle$ state to
have a long lifetime as shown in Fig. \ref{fig:PL}.

\section{Conclusions}

\label{sec:Conclusion}

We presented a detailed study of the structural, electronic, and optical
properties of TQDs of increasing sizes using both the first-principles
DFT as well as a CI approach based on the PPP model. The first-principles
geometry optimization revealed that all the TQDs considered in this
work are strictly planar with negative binding energies indicating
that they are thermodynamically stable. Furthermore, their vibrational
frequencies were found to be real, suggesting that the TQDs are dynamically
stable as well. Therefore, it should be possible to synthesize them
in laboratory.

The planar structure of TQDs implies that these are $\pi$-conjugated
molecules similar to benzoid hydrocarbons except that they consist
of 4- and 8-membered rings, without any hexagons. With the aim of
parameterizing an effective $\pi$-electron Hamiltonian , i.e., the
PPP-model for TQDs, we performed TDDFT calculations of their optical
absorption spectra, and compared it to that computed using the PPP-CI
approach, and found that the two sets of results agree well for the
screened Coulomb parameters of the PPP model.

By considering the symmetries of the low-lying excited states, and
corresponding transition dipoles, we demonstrated that TQDs will exhibit
photoluminescence, indicating their possible utility in optoelectronics.

Spin gap of the TQDs was computed predicting it to be much smaller
than the optical gap. This suggests an interesting possibility of
singlet fission in these systems into several triplets hinting their
utility in photovoltaics as well. Furthermore, by careful analysis
of the triplet state wave functions it was found that the lowest triplet
state $T_{1}$ exhibits significant configuration mixing caused by
electron-correlation effects.

The fact that TQDs are centro-symmetric implies that the linear optical
spectra computed in this work allows us to understand the nature of
the excited states with opposite inversion symmetry compared to the
ground state. However, to understand the states of the same inversion
symmetry as ground state we need to compute the nonlinear optical
absorption of these systems. The lowest-order non-linear optical response
of TQDs will be in the third order corresponding to processes such
as two-photon absorption and third-harmonic generation. The calculations
of nonlinear and excited state absorption spectra of the TQDs are
currently in progress in our group.

\section*{Conflicts of interest}

There are no conflicts to declare.

\section*{Data availability}

The computational details and data supporting this article have been
included as part of the main manuscript and the supplementary information.

\section*{Acknowledgment }

A.N. would like to acknowledge financial assistance from Indian Institute
of Technology Bombay under the teaching assistant (TA) category.

 \bibliographystyle{elsarticle-num}
\bibliography{references_revised}

\end{document}


\title{Supporting Information\\
 Computational study of geometry, electronic structure, and low-lying
excited states of linear T-graphene quantum dots}
\author{Arifa Nazir}
\email{arifabhatt9@gmail.com}

\affiliation{Department of Physics, Indian Institute of Technology Bombay, Powai,
Mumbai 400076 India}
\author{Alok Shukla}
\email{shukla@iitb.ac.in}

\affiliation{Department of Physics, Indian Institute of Technology Bombay, Powai,
Mumbai 400076 India}
\maketitle

\section{Cohesive energy per atom calculated for hydrocarbons compared with
that of TQDs}

\begin{figure}[H]
\begin{centering}
\includegraphics[scale=3]{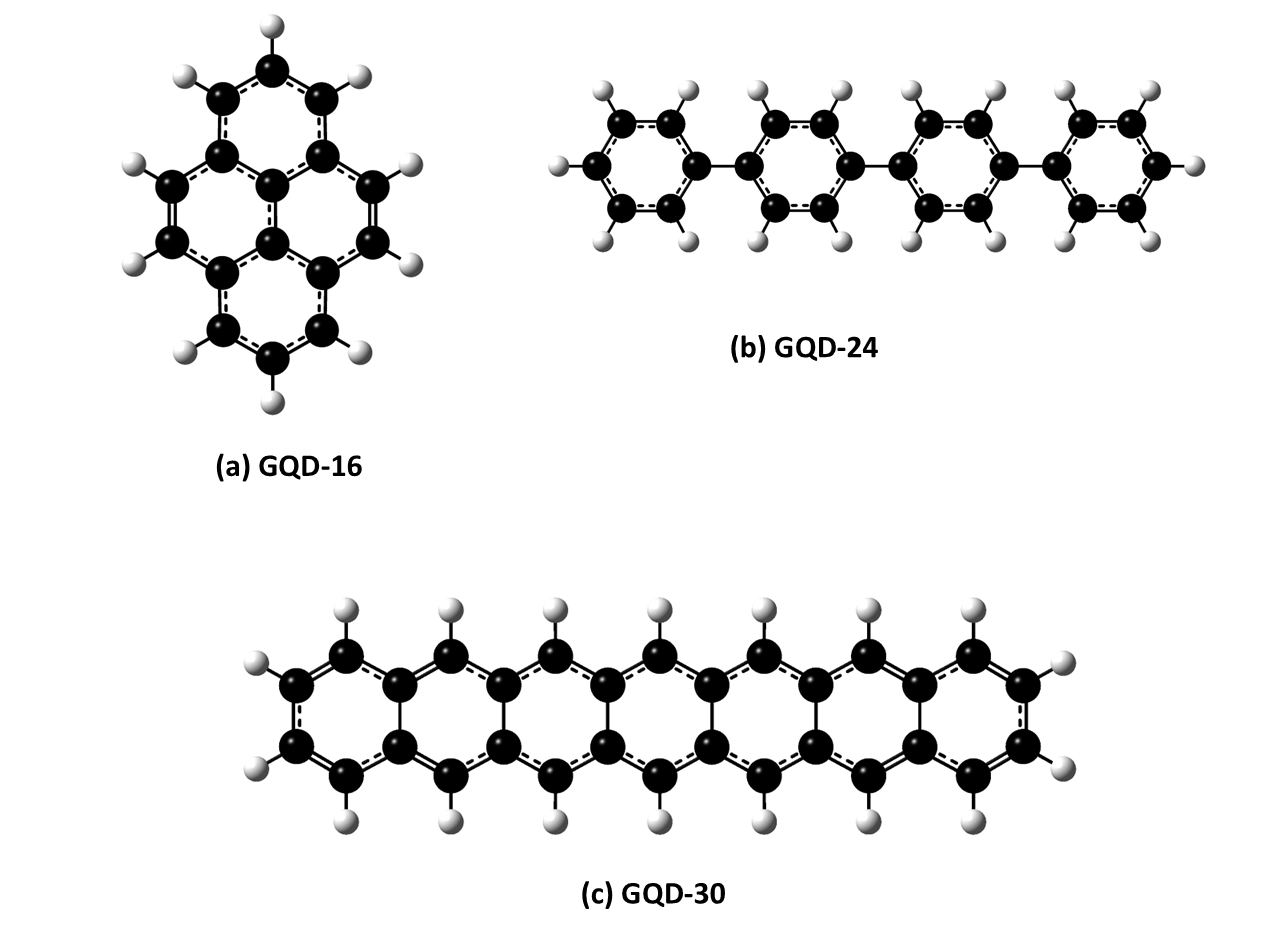} 
\par\end{centering}
\centering{}\caption{Three hexagonal graphene quantum dots (GQDs) of different shapes and
sizes considered for computing the cohesive energies per atom, for
comparison with the corresponding quantity of TQDs of similar sizes.
GQD-16 is pyrene, GQD-24 is p-quaterphenyl, and GQD-30 is heptacene.}
\end{figure}

\begin{table}[H]
\centering{}%
\begin{tabular}{ccc}
\toprule 
System  & chemical formula  & \begin{cellvarwidth}[t]
\centering
 Cohesive energy

per atom (eV) 
\end{cellvarwidth}\tabularnewline
\midrule
\midrule 
GQD-16  & C$_{16}$H$_{10}$  & -6.0718\tabularnewline
\midrule 
GQD-24  & C$_{24}$H$_{18}$  & -5.7226\tabularnewline
\midrule 
GQD-30  & C$_{30}$H$_{18}$  & -6.1081\tabularnewline
\bottomrule
\end{tabular}\caption{Cohesive energy per atom calculated for GQDs of different shapes and
sizes.}
\end{table}

\section{Detailed Information About PPP-CI Optical Absorption Spectrum}

In the provided tables, we offer detailed information on excitation
energies, predominant many-body wave-functions, and transition dipole
matrix elements for excited states relative to the ground state (1$^{1}$A$_{g}$).
The charge-conjugated configuration of a given configuration is abbreviated
as '$c.c.$'. The sign (+/-) preceding '$c.c.$' indicates whether
the two coefficients share the same or opposite signs. The notation
$H$ represents the Highest Occupied Molecular Orbital (HOMO), while
$'L'$ represents the Lowest Unoccupied Molecular Orbital (LUMO).
Similarly, $H-n$ and $L+m$ denote the $n$-th orbital below HOMO,
and the m-th orbital above LUMO, respectively.

The symbol $\ket{H\rightarrow L}$ denotes a singly excited configuration
achieved by promoting one electron from HOMO to LUMO, from the closed-shell
Hartree-Fock reference state. Similarly, the meanings of other singly-excited
configurations, such as $\ket{H-1\rightarrow L+1}$, can be deduced.
The symbol $\ket{H\rightarrow L+1;H-1\rightarrow L+2}$ denotes a
doubly excited configuration obtained by exciting two electrons from
the Hartree-Fock reference state: one from HOMO to LUMO+1, and the
other from HOMO-1 to LUMO+2. Excitation of three electrons can be
denoted in a similar manner e.g., $\ket{H\rightarrow L+1;H-1\rightarrow L+2;H\rightarrow L+2}$.

\begin{table}[H]
\protect\caption{Excited states giving rise to peaks in the singlet linear optical
absorption spectrum of TQD-16, computed employing the MRSDCI approach,
along with the screened parameters in the PPP model Hamiltonian. }
\label{wave_analysis_16_atoms_scr_TGQD}

\centering{}%
\begin{tabular}{>{\centering}p{1.5cm}>{\centering}p{1.5cm}>{\centering}p{1.5cm}>{\centering}p{1.5cm}>{\raggedright}p{6cm}}
\toprule 
Peak  & State  & E(eV)  & Transition dipole$(\text{\AA})$  & Dominant configurations, with their coefficients inside parentheses\tabularnewline
\midrule 
$DF_{x}^{(1)}$  & $1^{1}B_{3u}$  & $1.66$  & $0.00$  & $\ket{H\rightarrow L+1}-c.c(0.57746)$\tabularnewline
\multirow{2}{1.5cm}{$DF_{y}^{(2)}$} & \multirow{2}{1.5cm}{$1^{1}B_{2u}$} & \multirow{2}{1.5cm}{$2.72$} & \multirow{2}{1.5cm}{$0.00$} & $\ket{H\rightarrow L;H-1\rightarrow L}-c.c(0.45062)$\tabularnewline
 &  &  &  & $\ket{H\rightarrow L+2}-c.c(0.34909)$\tabularnewline
\multirow{3}{1.5cm}{$DF_{y}^{(3)}$} & \multirow{3}{1.5cm}{$2^{1}B_{2u}$} & \multirow{3}{1.5cm}{$3.38$} & \multirow{3}{1.5cm}{$0.022$} & $\ket{H\rightarrow L;H-1\rightarrow L}+c.c(0.5436)$\tabularnewline
 &  &  &  & $\ket{H-1\rightarrow L;H-1\rightarrow L+1}+c.c(0.2949)$\tabularnewline
 &  &  &  & $\ket{H\rightarrow L+1;H-2\rightarrow L;H-1\rightarrow L}-c.c(0.1062)$\tabularnewline
\multirow{2}{1.5cm}{$I_{x}^{(MI)}$} & \multirow{2}{1.5cm}{$2^{1}B_{3u}$} & \multirow{2}{1.5cm}{$3.52$} & \multirow{2}{1.5cm}{$-2.3280$} & $\ket{H\rightarrow L+1}+c.c(0.5881)$\tabularnewline
 &  &  &  & $\ket{H\rightarrow L;H\rightarrow L+1;H-1\rightarrow L+1}+c.c(0.2218)$\tabularnewline
\multirow{3}{1.5cm}{$II_{y}$} & \multirow{3}{1.5cm}{$5^{1}B_{2u}$} & \multirow{3}{1.5cm}{$4.50$} & \multirow{3}{1.5cm}{$0.723$} & $\ket{H\rightarrow L+1;H-1\rightarrow L+1}+c.c(0.4196)$\tabularnewline
 &  &  &  & $\ket{H\rightarrow L+2}+c.c(0.3728)$\tabularnewline
 &  &  &  & $\ket{H\rightarrow L+1;H\rightarrow L}+c.c(0.2203)$\tabularnewline
\multirow{3}{1.5cm}{$III_{y}$} & \multirow{3}{1.5cm}{$7^{1}B_{2u}$} & \multirow{3}{1.5cm}{$4.97$} & \multirow{3}{1.5cm}{$0.8286$} & $\ket{H\rightarrow L+2}+c.c(0.4485)$\tabularnewline
 &  &  &  & $\ket{H-1\rightarrow L+1;H-1\rightarrow L}+c.c(0.3317)$\tabularnewline
 &  &  &  & $\ket{H\rightarrow L+2;H-1\rightarrow L+1;H\rightarrow L}-c.c(0.1806)$\tabularnewline
\multirow{5}{1.5cm}{$IV_{x}$} & \multirow{2}{1.5cm}{$8^{1}B_{3u}$} & \multirow{2}{1.5cm}{$5.56$} & \multirow{2}{1.5cm}{$0.6718$} & $\ket{H\rightarrow L;H-2\rightarrow L}+c.c(0.5138)$\tabularnewline
 &  &  &  & $\ket{H-2\rightarrow L+1;H-1\rightarrow L}-c.c(0.2630)$\tabularnewline
 & \multirow{3}{1.5cm}{$9^{1}B_{3u}$} & \multirow{3}{1.5cm}{$5.60$} & \multirow{3}{1.5cm}{$-0.4551$} & $\ket{H-4\rightarrow L}+c.c(0.4163)$\tabularnewline
 &  &  &  & $\ket{H-1\rightarrow L;H-3\rightarrow L}-c.c(0.2159)$\tabularnewline
 &  &  &  & $\ket{H-2\rightarrow L+1;H-1\rightarrow L}-c.c(0.2159)$\tabularnewline
\multirow{2}{1.5cm}{$V_{y}$} & \multirow{2}{1.5cm}{$11^{1}B_{2u}$} & \multirow{2}{1.5cm}{$5.88$} & \multirow{2}{1.5cm}{$-1.1352$} & $\ket{H-1\rightarrow L+3}-c.c(0.5580)$\tabularnewline
 &  &  &  & $\ket{H\rightarrow L+2;H-4\rightarrow L+2}+c.c(0.1581)$\tabularnewline
\multirow{2}{1.5cm}{$VI_{y}$} & \multirow{2}{1.5cm}{$14^{1}B_{2u}$} & \multirow{2}{1.5cm}{$6.37$} & \multirow{2}{1.5cm}{$0.3784$} & $\ket{H\rightarrow L+2}+c.c(0.3793)$\tabularnewline
 &  &  &  & $\ket{H-5\rightarrow L;H-1\rightarrow L}+c.c(0.2806)$\tabularnewline
\multirow{2}{1.5cm}{$VII_{x}$} & \multirow{2}{1.5cm}{$17^{1}B_{3u}$} & \multirow{2}{1.5cm}{$6.8$} & \multirow{2}{1.5cm}{$0.3989$} & $\ket{H\rightarrow L+3;H\rightarrow L+1}-c.c(0.3771)$\tabularnewline
 &  &  &  & $\ket{H-2\rightarrow L+1;H-1\rightarrow L}+c.c(0.3467)$\tabularnewline
\multirow{2}{1.5cm}{$VIII_{x}$} & \multirow{2}{1.5cm}{$21^{1}B_{3u}$} & \multirow{2}{1.5cm}{$7.13$} & \multirow{2}{1.5cm}{$0.2520$} & $\ket{H-1\rightarrow L+3;H-1\rightarrow L+1}+c.c(0.2831)$\tabularnewline
 &  &  &  & $\ket{H\rightarrow L+1;H-3\rightarrow L}+c.c(0.2674)$\tabularnewline
\bottomrule
\end{tabular}
\end{table}

\begin{table}[H]
\protect\caption{Excited states giving rise to peaks in the singlet linear optical
absorption spectrum of TQD-16, computed employing the MRSDCI approach,
along with the standard parameters in the PPP model Hamiltonian. }
\label{wave_analysis_16atoms_std_TGQD}

\centering{}{\small{}{}}{\small{}%
\begin{tabular}{>{\raggedright}m{1.5cm}>{\raggedright}m{1.5cm}>{\raggedright}m{1.5cm}>{\raggedright}m{1.5cm}>{\raggedright}p{6.5cm}}
\hline 
{\small Peak  } & {\small State  } & {\small E(eV)  } & {\small Transition dipole$\text{(\AA)}$  } & {\small Dominant configurations, with their coefficients inside parentheses}\tabularnewline
\hline 
\multirow{1}{1.5cm}{{\small$DF_{x}^{(1)}$}} & \multirow{1}{1.5cm}{{\small$1^{1}B_{3u}$}} & \multirow{1}{1.5cm}{{\small$1.38$}} & \multirow{1}{1.5cm}{{\small$0.00$}} & {\small$\ket{H-1\rightarrow L}+c.c(0.5569)$}{\small\par}

{\small$\ket{H-2\rightarrow L+1;H-1\rightarrow L}-c.c(0.2441)$}{\small\par}

{\small$\ket{H\rightarrow L+1;H\rightarrow L+1;H-1\rightarrow L}+c.c(0.1313)$ }\tabularnewline
{\small$DF_{y}^{(2)}$  } & {\small$1^{1}B_{2u}$  } & {\small$2.58$  } & {\small$0.00$  } & {\small$\ket{H\rightarrow L+1;H\rightarrow L}+c.c(0.43433)$}{\small\par}

{\small$\ket{H\rightarrow L+2}+c.c(0.3535)$}{\small\par}

{\small$\ket{H\rightarrow L+1;H-2\rightarrow L+2}-c.c(0.1634)$}\tabularnewline
{\small$DF_{y}^{(3)}$  } & {\small$2^{1}B_{2u}$  } & {\small$2.79$  } & {\small$0.0320$  } & {\small$\ket{H\rightarrow L+1;H\rightarrow L}-c.c(0.50547)$}{\small\par}

{\small$\ket{H-1\rightarrow L+1;H-1\rightarrow L}-c.c(0.27485)$}{\small\par}

{\small$\ket{H\rightarrow L+1;H\rightarrow L;H-1\rightarrow L}-c.c(0.2055)$}\tabularnewline
{\small$I_{x}^{(MI)}$  } & {\small$3^{1}B_{3u}$  } & {\small$3.98$  } & {\small$-2.2628$  } & {\small$\ket{H-1\rightarrow L}-c.c(0.5627)$}{\small\par}

{\small$\ket{H\rightarrow L+1;H\rightarrow L+1;H-1\rightarrow L}-c.c(0.1313)$}\tabularnewline
{\small$II_{y}$  } & {\small$5^{1}B_{2u}$  } & {\small$4.61$  } & {\small$-0.8199$  } & {\small$\ket{H\rightarrow L+2}-c.c(0.5163)$}{\small\par}

{\small$\ket{H-1\rightarrow L+3}+c.c(0.2265)$}\tabularnewline
{\small$III_{x}$  } & {\small$7^{1}B_{3u}$  } & {\small$5.36$  } & {\small$0.6899$  } & {\small$\ket{H-4\rightarrow L}-c.c(0.3467)$}{\small\par}

{\small$\ket{H\rightarrow L+1;H-1\rightarrow L+2}+c.c(0.2627)$}{\small\par}

{\small$\ket{H-1\rightarrow L;H-3\rightarrow L}+c.c(0.2524)$}{\small\par}

{\small$\ket{H\rightarrow L+3;H-1\rightarrow L}-c.c(0.222)$ }\tabularnewline
{\small$IV_{y}$  } & {\small$12^{1}B_{2u}$  } & {\small$5.97$  } & {\small$-1.1772$  } & {\small$\ket{H-1\rightarrow L+3}+c.c(0.4273)$}{\small\par}

{\small$\ket{H\rightarrow L+2}-c.c(0.2567)$}{\small\par}

{\small$\ket{H\rightarrow L+1;H-4\rightarrow L+1}-c.c(0.14171)$}\tabularnewline
{\small$V_{y}$  } & {\small$16^{1}B_{2u}$  } & {\small$6.53$  } & {\small$-0.5946$  } & {\small$\ket{H-4\rightarrow L+1;H-1\rightarrow L}$ or}{\small\par}

{\small$\ket{H-4\rightarrow L;H-1\rightarrow L+1}+c.c(0.2614)$}{\small\par}

{\small$\ket{H\rightarrow L;H-1\rightarrow L;H-1\rightarrow L+2}-c.c(0.18736)$
or}{\small\par}

{\small$\ket{H\rightarrow L+2;H-1\rightarrow L;H-1\rightarrow L}-c.c(0.18736)$}\tabularnewline
{\small$VI_{y}$  } & {\small$20^{1}B_{2u}$  } & {\small$7.18$  } & {\small$-0.5608$  } & {\small$\ket{H\rightarrow L+2}+c.c(0.2687)$}{\small\par}

{\small$\ket{H\rightarrow L+1;H-2\rightarrow L+2}$}{\small\par}

{\small$\ket{H-1\rightarrow L+3}+c.c(0.2060)$}\tabularnewline
{\small$VII_{x}$  } & {\small$30^{1}B_{3u}$  } & {\small$7.78$  } & {\small$0.2714$  } & {\small$\ket{H\rightarrow L;H-6\rightarrow L}+c.c(0.3092)$}{\small\par}

{\small$\ket{H-2\rightarrow L+3}+c.c(0.2176)$}\tabularnewline
{\small$VIII_{y}$  } & {\small$30^{1}B_{2u}$  } & {\small$7.87$  } & {\small$0.2696$  } & {\small$\ket{H\rightarrow L+2}+c.c(0.2738)$}\tabularnewline
\hline 
\end{tabular}}{\small\par}
\end{table}

\begin{table}[H]
\protect\caption{Excited states giving rise to peaks in the singlet linear optical
absorption spectrum of TQD-24, computed employing the MRSDCI approach,
along with the screened parameters in the PPP model Hamiltonian.}
\label{wave_analysis_24atoms_scr_TGQD-1}

\centering{}%
\begin{tabular}{>{\raggedright}m{1.5cm}>{\raggedright}m{1.5cm}>{\raggedright}m{1.5cm}>{\raggedright}m{1.5cm}>{\raggedright}p{6.5cm}}
\toprule 
Peak  & State  & E(eV)  & Transition dipole$(\text{\AA})$  & Dominant configurations, with their coefficients inside parentheses\tabularnewline
\midrule 
\multirow{1}{1.5cm}{$DF_{x}$} & \multirow{1}{1.5cm}{$1^{1}B_{3u}$} & \multirow{1}{1.5cm}{$1.36$} & \multirow{1}{1.5cm}{$0.00$} & $\ket{H\rightarrow L+1}+c.c(0.5521)$

$\ket{H\rightarrow L+1;H-2\rightarrow L}-c.c(0.1854)$\tabularnewline
$I_{y}$  & $3^{1}B_{2u}$  & $2.70$  & $-0.2004$  & $\ket{H\rightarrow L+1;H\rightarrow L}-c.c(0.4390)$

$\ket{H\rightarrow L+1;H-1\rightarrow L+1}-c.c(0.2972)$

$\ket{H-2\rightarrow L+1}-c.c(0.2076)$\tabularnewline
$II_{x}^{(MI)}$  & $3^{1}B_{3u}$  & $2.79$  & $-3.2253$  & $\ket{H\rightarrow L+1}-c.c(0.56742)$

$\ket{H\rightarrow L+1;H\rightarrow L+1;H-1\rightarrow L}-c.c(0.1824)$\tabularnewline
$III_{y}$  & $8^{1}B_{2u}$  & $3.988$  & $1.3034$  & $\ket{H\rightarrow L+3}+c.c(0.5325)$

$\ket{H\rightarrow L+2;H-1\rightarrow L;H\rightarrow L}-c.c(0.1314)$

$\ket{H-2\rightarrow L;H-2\rightarrow L+1}-c.c(0.1232)$\tabularnewline
$IV_{x}$  & $11^{1}B_{3u}$  & $4.35$  & $-0.6805$  & $\ket{H\rightarrow L+1;H\rightarrow L+2}-c.c(0.3900)$

$\ket{H-1\rightarrow L+1;H-1\rightarrow L+1}-c.c(0.2493)$

$\ket{H\rightarrow L+1;H-2\rightarrow L}+c.c(0.1904)$\tabularnewline
$V_{y}$  & $11^{1}B_{2u}$  & $4.56$  & $-0.2748$  & $\ket{H\rightarrow L+2;H-1\rightarrow L+2}-c.c(0.4105)$

$\ket{H-1\rightarrow L;H-2\rightarrow L+2}+c.c(0.1851)$\tabularnewline
$VI_{x}$  & $16^{1}B_{3u}$  & $4.92$  & $-0.5939$  & $\ket{H\rightarrow L+3;H\rightarrow L}+c.c(0.4285)$

$\ket{H-3\rightarrow L+1;H-1\rightarrow L}-c.c(0.2872)$

$\ket{H-5\rightarrow L}-c.c(0.1205)$\tabularnewline
$VII_{x}$  & $22^{1}B_{3u}$  & $5.34$  & $-0.7607$  & $\ket{H-3\rightarrow L+1;H-1\rightarrow L}-c.c(0.3494)$

$\ket{H-1\rightarrow L+4}+c.c(0.2672)$\tabularnewline
$VIII_{x}$  & $24^{1}B_{3u}$  & $5.65$  & $0.5378$  & $\ket{H-5\rightarrow L}-c.c(0.3020)$

$\ket{H\rightarrow L+7}+c.c(0.2414)$

$\ket{H-3\rightarrow L+1;H-1\rightarrow L}-c.c(0.2171)$\tabularnewline
$IX_{x}$  & $28^{1}B_{3u}$  & $5.8$8  & $0.3864$  & $\ket{H\rightarrow L+7}+c.c(0.3223)$

$\ket{H-1\rightarrow L+1;H-1\rightarrow L}+c.c(0.2437)$

$\ket{H-1\rightarrow L+4}+c.c(0.23395)$\tabularnewline
\bottomrule
\end{tabular}
\end{table}

\begin{table}[H]
\protect\caption{Excited states giving rise to peaks in the singlet linear optical
absorption spectrum of TQD-24, computed employing the MRSDCI approach,
along with the standard parameters in the PPP model Hamiltonian.}
\label{wave_analysis_24atoms_std_TGQD-1}

\centering{}%
\begin{tabular}{>{\raggedright}m{1.5cm}>{\raggedright}m{1.5cm}>{\raggedright}m{1.5cm}>{\raggedright}m{1.5cm}>{\raggedright}p{6.5cm}}
\hline 
Peak  & State  & E(eV)  & Transition dipole$(\text{\AA})$  & Dominant configurations, with their coefficients inside parentheses\tabularnewline
\hline 
\multirow{2}{1.5cm}{$DF_{x}$} & \multirow{2}{1.5cm}{$1^{1}B_{3u}$} & \multirow{2}{1.5cm}{$1.16$} & \multirow{2}{1.5cm}{$0.00$} & $\ket{H\rightarrow L+1}+c.c(0.4885)$\tabularnewline
 &  &  &  & $\ket{H\rightarrow L;H-2\rightarrow L+1}+c.c(0.3062)$

$\ket{H-2\rightarrow L+3}+c.c(0.1173)$\tabularnewline
$I_{y}$  & $2^{1}B_{2u}$  & $2.35$  & $-0.1670$  & \multirow{3}{6.5cm}{$\ket{H-2\rightarrow L+1}+c.c(0.4743)$

$\ket{H\rightarrow L;H\rightarrow L+1}-c.c(0.2521)$

$\ket{H\rightarrow L;H\rightarrow L;H-1\rightarrow L+2}+c.c(0.1782)$}\tabularnewline
 &  &  &  & \tabularnewline
 &  &  &  & \tabularnewline
$II_{x}^{(MI)}$  & $4^{1}B_{3u}$  & $3.22$  & $3.1179$  & $\ket{H\rightarrow L+1}-c.c(0.5616)$

$\ket{H\rightarrow L+1;H\rightarrow L+1;H-1\rightarrow L}-c.c(0.1764)$\tabularnewline
\multirow{1}{1.5cm}{} & \begin{turn}{90}
\end{turn} &  &  & $\ket{H-3\rightarrow L;H-1\rightarrow L+1}-c.c(0.1122)$\tabularnewline
$III_{y}$  & $6^{1}B_{2u}$  & $3.96$  & $0.8411$  & $\ket{H\rightarrow L+3}+c.c(0.4150)$

$\ket{H\rightarrow L+2;H-1\rightarrow L+2}-c.c(0.2167)$

$\ket{H-2\rightarrow L+1;H\rightarrow L+2}+c.c(0.1631)$\tabularnewline
$IV_{x}$  & $10^{1}B_{3u}$  & $4.25$  & $-0.36177$  & $\ket{H\rightarrow L+1;H-1\rightarrow L;H-1\rightarrow L}-c.c(0.3385)$

$\ket{H-2\rightarrow L+3}-c.c(0.2154)$

$\ket{H\rightarrow L;H-1\rightarrow L+2}-c.c(0.1960)$\tabularnewline
$V_{x}$  & $12^{1}B_{3u}$  & $4.50$  & $0.44520$  & $\ket{H\rightarrow L;H-1\rightarrow L+2}-c.c(0.3007)$

$\ket{H\rightarrow L+1;H-3\rightarrow L+1}+c.c(0.2429)$

$\ket{H-4\rightarrow L+1}-c.c(0.2072)$\tabularnewline
$VI_{x}$  & $16^{1}B_{3u}$  & $5.10$  & $0.4904$  & $\ket{H\rightarrow L;H\rightarrow L+3}+c.c(0.4906)$

$\ket{H-3\rightarrow L+1;H-3\rightarrow L}+c.c(0.1417)$\tabularnewline
\multirow{1}{1.5cm}{} &  &  &  & $\ket{H\rightarrow L;H\rightarrow L;H-3\rightarrow L+2}-c.c(0.1245)$\tabularnewline
$VII_{x}$  & $21^{1}B_{3u}$  & $5.50$  & $-0.5834$  & $\ket{H\rightarrow L+7}-c.c(0.2011)$

$\ket{H\rightarrow L+2;H-1\rightarrow L}-c.c(0.1972)$

$\ket{H-2\rightarrow L;H\rightarrow L+3}+c.c(0.1721)$\tabularnewline
\multirow{2}{1.5cm}{$VIII_{x\&y}$} & $28^{1}B_{2u}$  & $5.75$  & $-0.3814$  & $\ket{H\rightarrow L;H-1\rightarrow L+4}-c.c(0.2073)$

$\ket{H\rightarrow L;H-2\rightarrow L+3}+c.c(0.1745)$\tabularnewline
 & $27^{1}B_{3u}$  & $5.75$  & $-0.18003$  & $\ket{H-1\rightarrow L+1;H-3\rightarrow L}-c.c(0.3147)$

$\ket{H-1\rightarrow L;H\rightarrow L+2}-c.c(0.2129)$\tabularnewline
\hline 
\end{tabular}
\end{table}

\begin{table}[H]
\protect\caption{Excited states giving rise to peaks in the singlet linear optical
absorption spectrum of TQD-32, computed employing the MRSDCI approach,
along with the screened parameters in the PPP model. }
\label{wave_analysis_32atoms_scr_TGQD}

\centering{}%
\begin{tabular}{>{\raggedright}m{1.5cm}>{\raggedright}m{1.5cm}>{\raggedright}m{1.5cm}>{\raggedright}m{1.5cm}>{\raggedright}p{6.5cm}}
\toprule 
Peak  & State  & E(eV)  & Transition dipole$(\text{\AA})$  & Dominant configurations, with their coefficients inside parentheses\tabularnewline
\midrule 
$DF_{x}$  & $1^{1}B_{3u}$  & $1.34$  & $0.00$  & $\ket{H\rightarrow L+1}+c.c(0.5053)$

$\ket{H\rightarrow L;H\rightarrow L+2}-c.c(0.2100)$\tabularnewline
$I_{y}$  & $2^{1}B_{2u}$  & $1.92$  & $0.3256$  & $\ket{H\rightarrow L+2}+c.c(0.5574)$

$\ket{H\rightarrow L+2;H\rightarrow L+2;H-2\rightarrow L}+c.c(0.1692)$\tabularnewline
$II_{x}$  & $4^{1}B_{3u}$  & $2.52$  & $-3.7016$  & $\ket{H\rightarrow L+1}-c.c(0.5435)$

$\ket{H\rightarrow L+1;H\rightarrow L+1;H-1\rightarrow L}-c.c(0.1276)$\tabularnewline
$III_{x}$  & $6^{1}B_{3u}$  & $3.05$  & $-0.5557$  & $\ket{H\rightarrow L;H-2\rightarrow L}+c.c(0.4253)$

$\ket{H-2\rightarrow L+2;H\rightarrow L+2}+c.c(0.3386)$

$\ket{H\rightarrow L+1}-c.c(0.1527)$

$\ket{H\rightarrow L+1;H-1\rightarrow L+2}-c.c(0.1291)$\tabularnewline
\multirow{3}{1.5cm}{$IV_{x}$} & \multirow{3}{1.5cm}{$10^{1}B_{3u}$} & \multirow{3}{1.5cm}{$3.60$} & \multirow{3}{1.5cm}{$1.3485$} & \multirow{3}{6.5cm}{$\ket{H-3\rightarrow L+2}+c.c(0.3582)$

$\ket{H\rightarrow L;H-1\rightarrow L+3}+c.c(0.2754)$

$\ket{H-1\rightarrow L+2;H\rightarrow L+1}-c.c(0.2136)$}\tabularnewline
 &  &  &  & \tabularnewline
 &  &  &  & \tabularnewline
$V_{x\&y}$  & $16^{1}B_{3u}$  & $4.06$  & $0.1227$  & $\ket{H-2\rightarrow L;H-1\rightarrow L;H\rightarrow L+2}-c.c(0.3706)$

$\ket{H-1\rightarrow L+1;H\rightarrow L+2}+c.c(0.2301)$

$\ket{H-1\rightarrow L+4}-c.c(0.2112)$\tabularnewline
 & $19^{1}B_{2u}$  & $4.06$  & $0.3777$  & $\ket{H-4\rightarrow L+2}-c.c(0.3039)$

$\ket{H-1\rightarrow L+1;H-1\rightarrow L}-c.c(0.2867)$

$\ket{H\rightarrow L;H-2\rightarrow L+3}-c.c(0.2378)$\tabularnewline
$VI_{x}$  & $19^{1}B_{3u}$  & $4.24$  & $-0.4362$  & $\ket{H\rightarrow L+1;H-1\rightarrow L+2}-c.c(0.3352)$

$\ket{H-4\rightarrow L+4;H\rightarrow L+2}+c.c(0.2063)$

$\ket{H-2\rightarrow L+2;H-1\rightarrow L,H\rightarrow L}+c.c(0.2030)$\tabularnewline
$VII_{x\&y}$  & $25^{1}B_{3u}$  & $4.69$  & $-0.4323$  & $\ket{H-4\rightarrow L+4;H\rightarrow L+2}+c.c(0.2817)$

$\ket{H-1\rightarrow L+1;H\rightarrow L+2}-c.c(0.2469)$

$\ket{H-2\rightarrow L+1;H\rightarrow L+1}+c.c(0.2239)$

$\ket{H-1\rightarrow L+4}-c.c(0.2047)$\tabularnewline
 & $28^{1}B_{2u}$  & $4.69$  & $-1.3962$  & $\ket{H\rightarrow L+5}-c.c(0.3795)$

$\ket{H-4\rightarrow L+2}-c.c(0.3144)$

$\ket{H-4\rightarrow L;H\rightarrow L+1}-c.c(0.2123)$\tabularnewline
\bottomrule
\end{tabular}
\end{table}

\begin{table}[H]
\protect\caption{Excited states giving rise to peaks in the singlet linear optical
absorption spectrum of TQD-32, computed employing the MRSDCI approach,
along with the standard parameters in the PPP model. }
\label{wave_analysis_32atoms_std_TGQD}

\centering{}{\scriptsize{}{}}{\scriptsize{}%
\begin{tabular}{>{\raggedright}m{1.5cm}>{\raggedright}m{1.5cm}>{\raggedright}m{1.5cm}>{\raggedright}m{1.5cm}>{\raggedright}p{6.5cm}}
\toprule 
Peak  & State  & E(eV)  & Transition dipole$(\text{\AA})$  & Dominant configurations, with their coefficients inside parentheses\tabularnewline
\midrule 
\multirow{1}{1.5cm}{$DF_{x}$} & \multirow{1}{1.5cm}{$1^{1}B_{3u}$} & \multirow{1}{1.5cm}{$1.$19} & \multirow{1}{1.5cm}{$0.00$} & $\ket{H\rightarrow L+1}-c.c(0.4545)$

$\ket{H\rightarrow L+1;H-1\rightarrow L+2}-c.c(0.2822)$

$\ket{H-3\rightarrow L+2}+c.c(0.1651)$\tabularnewline
$I_{y}$  & $1^{1}B_{2u}$  & $1.59$  & $0.2533$  & $\ket{H\rightarrow L+2}-c.c(0.5506)$

$\ket{H-4\rightarrow L+2}-c.c(0.1407)$

$\ket{H-2\rightarrow L;H-2\rightarrow L+2;H\rightarrow L+2}-c.c(0.1145)$\tabularnewline
$II_{x}$  & $4^{1}B_{3u}$  & $2.96$  & $-3.5166$  & $\ket{H\rightarrow L+1}+c.c(0.5364)$

$\ket{H\rightarrow L+3;H-1\rightarrow L}-c.c(0.1420)$$\ket{H-1\rightarrow L;H-1\rightarrow L;H\rightarrow L+1}+c.c(0.1080)$\tabularnewline
\multirow{1}{1.5cm}{$III_{x}$} & $9B_{3u}$  & \multirow{1}{1.5cm}{$3.69$} & $0.4577$  & $\ket{H-2\rightarrow L;H-1\rightarrow L+1}+c.c(0.3052)$$\ket{H\rightarrow L+3;H-1\rightarrow L}-c.c(0.2774)$$\ket{H-2\rightarrow L+1;H-1\rightarrow L;H-1\rightarrow L+2}-c.c(0.1436)$\tabularnewline
\multirow{9}{1.5cm}{$IV_{x}$} & \multirow{9}{1.5cm}{$12B_{3u}$} & \multirow{9}{1.5cm}{$3.83$} & \multirow{9}{1.5cm}{$0.9072$} & \multirow{9}{6.5cm}{$\ket{H-2\rightarrow L+3}-c.c(0.2312)$$\ket{H\rightarrow L+3;H-1\rightarrow L}-c.c(0.2312)$$\ket{H-2\rightarrow L;H-1\rightarrow L+1}+c.c(0.1787)$$\ket{H-1\rightarrow L;H-1\rightarrow L;H\rightarrow L+1}+c.c(0.1705)$}\tabularnewline
 &  &  &  & \tabularnewline
 &  &  &  & \tabularnewline
 &  &  &  & \tabularnewline
 &  &  &  & \tabularnewline
 &  &  &  & \tabularnewline
 &  &  &  & \tabularnewline
 &  &  &  & \tabularnewline
 &  &  &  & \tabularnewline
$V_{x}$  & $16B_{3u}$  & $4.18$  & $-0.3402$  & $\ket{H\rightarrow L+1;H\rightarrow L;H-2\rightarrow L+2}-c.c(0.3787)$

$\ket{H-2\rightarrow L;H\rightarrow L+4}+c.c(0.2299)$

$\ket{H-2\rightarrow L+2;H-1\rightarrow L+4}+c.c(0.1437)$\tabularnewline
$VI_{y}$  & $17B_{2u}$  & $4.30$  & $-0.3402$  & $\ket{H-4\rightarrow L+2}+c.c(0.2660)$

$\ket{H-2\rightarrow L+3;H\rightarrow L}+c.c(0.2297)$

$\ket{H\rightarrow L+1;H\rightarrow L+2;H-1\rightarrow L}+c.c(0.1892)$\tabularnewline
$VII_{y}$  & $20B_{2u}$  & $4.68$  & $1.156484$  & $\ket{H-5\rightarrow L}+c.c(0.3899)$

$\ket{H-4\rightarrow L}-c.c(0.3209)$

$\ket{H\rightarrow L+2}-c.c(0.1175)$\tabularnewline
\bottomrule
\end{tabular}}{\scriptsize\par}
\end{table}

\begin{table}[H]
\protect\caption{Excited states giving rise to peaks in the singlet linear optical
absorption spectrum of TQD-40, computed employing the MRSDCI approach,
along with the screened parameters in the PPP model. }
\label{wave_analysis_40atoms_scr_TGQD}

\centering{}%
\begin{tabular}{>{\raggedright}m{1.5cm}>{\raggedright}m{1.5cm}>{\raggedright}m{1.5cm}>{\raggedright}m{1.5cm}>{\raggedright}p{6.5cm}}
\hline 
Peak  & State  & E(eV)  & Transition dipole$(\text{\AA})$  & Dominant configurations, with their coefficients inside parentheses\tabularnewline
\hline 
\multirow{1}{1.5cm}{$DF_{x}$} & \multirow{1}{1.5cm}{$1^{1}B_{3u}$} & \multirow{1}{1.5cm}{$1.25$} & \multirow{1}{1.5cm}{$0.00$} & $\ket{H\rightarrow L+1}+c.c(0.5030)$

$\ket{H-2\rightarrow L;H\rightarrow L}+c.c(0.1656)$

$\ket{H-2\rightarrow L+3}+c.c(0.1468)$\tabularnewline
$I_{y}$  & $2^{1}B_{2u}$  & $1.72$  & $-0.1841$  & $\ket{H\rightarrow L+2}-c.c(0.5696)$

$\ket{H-2\rightarrow L;H-2\rightarrow L;H\rightarrow L+2}-c.c(0.1454)$\tabularnewline
\multirow{1}{1.5cm}{$II_{x}$} & $5^{1}B_{3u}$  & $2.21$  & $-4.6760$  & $\ket{H\rightarrow L+1}-c.c(0.5677)$

$\ket{H-1\rightarrow L;H-1\rightarrow L;H\rightarrow L+1}-c.c(0.1062)$\tabularnewline
$III_{x}$  & $7^{1}B_{3u}$  & $2.79$  & $0.4978$  & $\ket{H\rightarrow L;H\rightarrow L+2}-c.c(0.3614)$

$\ket{H-2\rightarrow L;H-2\rightarrow L+2}-c.c(0.2860)$

$\ket{H\rightarrow L+1;H-1\rightarrow L+2}+c.c(0.2216)$\tabularnewline
 & $8^{1}B_{3u}$  & $2.83$  & $-1.2006$  & $\ket{H-4\rightarrow L}-c.c(0.3146)$

$\ket{H-1\rightarrow L;H-3\rightarrow L}-c.c(0.2291)$

$\ket{H\rightarrow L;H\rightarrow L+2}-c.c(0.2259)$

$\ket{H-3\rightarrow L+2}-c.c(0.2216)$\tabularnewline
$IV_{x}$  & $13^{1}B_{3u}$  & $3.17$  & $1.7899$  & $\ket{H-3\rightarrow L+2}-c.c(0.3219)$

$\ket{H-1\rightarrow L;H-1\rightarrow L+2}-c.c(0.2661)$

$\ket{H-5\rightarrow L+1}-c.c(0.2387)$\tabularnewline
$V_{x}$  & $21^{1}B_{3u}$  & $3.81$  & $0.6146$  & $\ket{H\rightarrow L+1;H-1\rightarrow L;H-1\rightarrow L}-c.c(0.35252)$

$\ket{H-5\rightarrow L+1}-c.c(0.2585)$

$\ket{H-2\rightarrow L;H-1\rightarrow L+1}+c.c(0.2212)$\tabularnewline
$VI_{x}$  & $25^{1}B_{3u}$  & $4.01$  & $0.7757$  & $\ket{H-2\rightarrow L;H-1\rightarrow L+1}+c.c(0.2122)$

$\ket{H-5\rightarrow L+1}-c.c(0.1995)$\tabularnewline
$VII_{y}$  & $25^{1}B_{2u}$  & $4.24$  & $1.9619$  & $\ket{H-6\rightarrow L}-c.c(0.3773)$

$\ket{H-5\rightarrow L+2}-c.c(0.3773)$\tabularnewline
\hline 
\end{tabular}
\end{table}

\begin{table}[H]
\protect\caption{Excited states giving rise to peaks in the singlet linear optical
absorption spectrum of TQD-40, computed employing the MRSDCI approach,
along with the standard parameters in the PPP model. }
\label{wave_analysis_40atoms_std_TGQD}

\centering{}%
\begin{tabular}{>{\raggedright}m{1.5cm}>{\raggedright}m{1.5cm}>{\raggedright}m{1.5cm}>{\raggedright}m{1.5cm}>{\raggedright}p{6.5cm}}
\hline 
Peak  & State  & E(eV)  & Transition dipole$(\text{\AA})$  & Dominant configurations, with their coefficients inside parentheses\tabularnewline
\hline 
\multirow{1}{1.5cm}{$DF_{x}$} & \multirow{1}{1.5cm}{$1^{1}B_{3u}$} & \multirow{1}{1.5cm}{$1.58$} & \multirow{1}{1.5cm}{$0.00$} & $\ket{H\rightarrow L+1}-c.c(0.4602)$

$\ket{H-3\rightarrow L+2}-c.c(0.2958)$\tabularnewline
$I_{y}$  & $1^{1}B_{2u}$  & $1.37$  & $-0.1509$  & $\ket{H\rightarrow L+2}+c.c(0.5463)$\tabularnewline
$II_{x}$  & $4^{1}B_{3u}$  & $2.82$  & $-4.1152$  & $\ket{H\rightarrow L+1}+c.c(0.5373)$

$\ket{H\rightarrow L+4}+c.c(0.1098)$\tabularnewline
\multirow{1}{1.5cm}{$III_{x}$} & $10^{1}B_{3u}$  & $3.75$  & $1.5615$  & $\ket{H-3\rightarrow L+2}+c.c(0.4082)$

$\ket{H-1\rightarrow L+5}+c.c(0.2036)$

$\ket{H-1\rightarrow L;H-1\rightarrow L+2}+c.c(0.1914)$\tabularnewline
$IV_{x\&y}$  & $22^{1}B_{2u}$  & $4.39$  & $-1.1715$  & $\ket{H-1\rightarrow L+1;H-3\rightarrow L+2}-c.c(0.2799)$

$\ket{H-2\rightarrow L+5}+c.c(0.2775)$

$\ket{H-6\rightarrow L}+c.c(0.2721)$\tabularnewline
 & $18^{1}B_{3u}$  & $4.42$  & $-0.4870$  & $\ket{H\rightarrow L+3;H-4\rightarrow L}-c.c(0.2953)$

$\ket{H\rightarrow L+3;H-3\rightarrow L+2}-c.c(0.2536)$

$\ket{H-2\rightarrow L+1;H-1\rightarrow L+1}-c.c(0.2090)$\tabularnewline
$V_{x}$  & $21^{1}B_{3u}$  & $4.80$  & $-0.6181$  & $\ket{H-2\rightarrow L+1;H-2\rightarrow L+3}+c.c(0.2291)$

$\ket{H-1\rightarrow L+5}+c.c(0.2241)$\tabularnewline
$VI_{x}$  & $23^{1}B_{3u}$  & $5.08$  & $0.61384$  & $\ket{H\rightarrow L+4}+c.c(0.2718)$

$\ket{H-1\rightarrow L+5}+c.c(0.2718)$\tabularnewline
\hline 
\end{tabular}
\end{table}

\section{Detailed Information About TDDFT Optical Absorption Spectrum}

\begin{table}[H]
\protect\caption{Excited states giving rise to peaks in the singlet linear optical
absorption spectrum of TQD-24 computed employing the TDDFT approach
at the $B3LYP/6311++G(d,p)$ level.}
\label{wave_analysis_24_tddft}

\centering{}%
\begin{tabular}{>{\raggedright}m{1.5cm}>{\raggedright}m{1.5cm}>{\raggedright}m{1.5cm}>{\raggedright}m{1.5cm}>{\raggedright}p{6.5cm}}
\hline 
Peak  & Symmetry  & E(eV)  & Oscillator strength  & Wavefunction\tabularnewline
\hline 
\multirow{1}{1.5cm}{$DF_{x}$} & \multirow{1}{1.5cm}{$1^{1}B_{3u}$} & \multirow{1}{1.5cm}{$1.4184$} & \multirow{1}{1.5cm}{$0.0471$} & $\ket{H-1\rightarrow L+1}(-0.42895)$\tabularnewline
 &  &  &  & $\ket{H\rightarrow L}(0.56344)$\tabularnewline
$I_{y}$  & $^{1}B_{2u}$  & $2.63$  & $0.0018$  & $\ket{H-2\rightarrow L+1}(0.27891)$

$\ket{H\rightarrow L+2}(0.64691)$\tabularnewline
\multirow{1}{1.5cm}{$II_{x}^{(MI)}$} & $^{1}B_{3u}$  & $2.9042$  & $2.9814$  & $\ket{H-1\rightarrow L+1}(0.57903)$

$\ket{H\rightarrow L}(0.46046)$\tabularnewline
$III_{y}$  & $^{1}B_{2u}$  & $4.0989$  & $0.0693$  & $\ket{H-3\rightarrow L+2}(0.59359)$

$\ket{H-2\rightarrow L+3}(0.31822)$\tabularnewline
$IV_{x}$  & $^{1}B_{3u}$  & $4.2489$  & $0.1696$  & $\ket{H-1\rightarrow L+3}(0.65840)$

$\ket{H-3\rightarrow L}(0.16318)$

$\ket{H-6\rightarrow L+1}(0.12060)$\tabularnewline
\hline 
\end{tabular}
\end{table}

\begin{table}[H]
\protect\caption{Excited states giving rise to peaks in the singlet linear optical
absorption spectrum of TQD-32, computed employing the TDDFT approach
at the $B3LYP/6311++G(d,p)$ level.}
\label{wave_analysis_32_tddft}

\centering{}%
\begin{tabular}{>{\raggedright}m{1.5cm}>{\raggedright}m{1.5cm}>{\raggedright}m{1.5cm}>{\raggedright}m{1.5cm}>{\raggedright}p{6.5cm}}
\hline 
Peak  & Symmetry  & E(eV)  & Oscillator strength  & Wavefunction\tabularnewline
\hline 
\multirow{1}{1.5cm}{$DF_{x}$} & \multirow{1}{1.5cm}{$^{1}B_{3u}$} & \multirow{1}{1.5cm}{$1.1573$} & \multirow{1}{1.5cm}{$0.0096$} & $\ket{H-1\rightarrow L}(0.53120)$\tabularnewline
 &  &  &  & $\ket{H\rightarrow L+2}(-0.46131)$\tabularnewline
$I_{y}$  & $^{1}B_{2u}$  & $1.5780$  & $0.0001$  & $\ket{H-2\rightarrow L}(0.67565)$

$\ket{H\rightarrow L+1}(-0.19779)$\tabularnewline
$II_{x}^{(MI)}$  & $^{1}B_{3u}$  & $2.3657$  & $3.5359$  & $\ket{H\rightarrow L+2}(0.54133)$\tabularnewline
 &  &  &  & $\ket{H-1\rightarrow L}(0.48934)$\tabularnewline
 &  &  &  & $\ket{H-1\rightarrow L+1}(0.10077)$\tabularnewline
$III_{x}$  & $^{1}B_{3u}$  & 2.6578  & 0.1813  & $\ket{H-1\rightarrow L+3}(0.65001)$

$\ket{H-3\rightarrow L+2}(-0.27122)$\tabularnewline
$IV_{y}$  & $^{1}B_{2u}$  & 3.2925  & 0.0327  & $\ket{H-4\rightarrow L+1}(0.68199)$\tabularnewline
\multirow{1}{1.5cm}{$V_{x}$} & $^{1}B_{3u}$  & 3.5966  & 0.4382  & $\ket{H-2\rightarrow L+3}(0.58251)$

$\ket{H-1\rightarrow L+4}(0.27947)$

$\ket{H-3\rightarrow L+1}(0.21123)$\tabularnewline
$VI_{x}$  & $^{1}B_{3u}$  & 4.1226  & 0.0623  & $\ket{H-1\rightarrow L+4}(0.53485)$

$\ket{H-6\rightarrow L}(-0.34651)$\tabularnewline
\hline 
\end{tabular}
\end{table}

\begin{table}[H]
\protect\caption{Excited states giving rise to peaks in the singlet linear optical
absorption spectrum of TQD-40, computed employing the TDDFT approach
at the $B3LYP/6311++G(d,p)$ level.}
\label{wave_analysis_40_tddft}

\centering{}%
\begin{tabular}{>{\raggedright}m{1.5cm}>{\raggedright}m{1.5cm}>{\raggedright}m{1.5cm}>{\raggedright}m{1.5cm}>{\raggedright}p{6.5cm}}
\hline 
Peak  & Symmetry  & E(eV)  & Oscillator strength  & Wavefunction\tabularnewline
\hline 
\multirow{1}{1.5cm}{$DF_{x}$} & \multirow{1}{1.5cm}{$^{1}B_{3u}$} & \multirow{1}{1.5cm}{$1.1038$} & \multirow{1}{1.5cm}{$0.0314$} & $\ket{H-1\rightarrow L}+(0.5396)$\tabularnewline
$I_{y}$  & $^{1}B_{2u}$  & $1.6488$  & $0.0009$  & $\ket{H-2\rightarrow L}(-0.66736)$

$\ket{H\rightarrow L+2}(-0.22128)$\tabularnewline
\multirow{3}{1.5cm}{$II_{x}^{(MI)}$} & \multirow{2}{1.5cm}{$^{1}B_{3u}$} & \multirow{2}{1.5cm}{$2.0938$} & \multirow{2}{1.5cm}{$3.3676$} & $\ket{H-4\rightarrow L}(-0.13142)$\tabularnewline
 &  &  &  & $\ket{H\rightarrow L+2}(0.47101)$

$\ket{H-1\rightarrow L}(0.44172)$

$\ket{H-3\rightarrow L+1}(0.26703)$

$\ket{H\rightarrow L+4}(0.12153)$\tabularnewline
 & $^{1}B_{3u}$  & 2.1529  & 1.1435  & $\ket{H-4\rightarrow L}(0.39334)$

$\ket{H-3\rightarrow L+1}(-0.33790)$

$\ket{H\rightarrow L+2}(0.28569)$

$\ket{H-2\rightarrow L}(0.23299)$

$\ket{H\rightarrow L+4}(-0.22716)$\tabularnewline
$III_{x}$  & $^{1}B_{3u}$  & $2.3596$  & $0.3018$  & $\ket{H-4\rightarrow L}(0.49923)$

$\ket{H-3\rightarrow L+1}(0.48155)$

$\ket{H\rightarrow L+2}(-0.11886)$\tabularnewline
$IV_{x}$  & $^{1}B_{3u}$  & 2.8528  & 0.2242  & $\ket{H\rightarrow L+2}(0.56963)$

$\ket{H-2\rightarrow L+3}(0.33995)$

$\ket{H-5\rightarrow L+2}(0.21577)$\tabularnewline
\multirow{2}{1.5cm}{$V_{x\&y}$} & $^{1}B_{3u}$  & 2.9913  & 0.0749  & $\ket{H-5\rightarrow L+2}(0.54184)$

$\ket{H-2\rightarrow L+3}(-0.43704)$\tabularnewline
 & $^{1}B_{2u}$  & 3.0650  & 0.0354  & $\ket{H-5\rightarrow L+1}(0.68415)$\tabularnewline
$VI_{x}$  & $^{1}B_{3u}$  & 3.3057  & 0.6161  & $\ket{H-1\rightarrow L+5}(0.51112)$

$\ket{H-2\rightarrow L+3}(0.25406)$

$\ket{H-5\rightarrow L+2}(0.22166)$

$\ket{H\rightarrow L+4}(-0.20149)$\tabularnewline
\hline 
\end{tabular}
\end{table}

\section{Comparison of PPP-CI and TDDFT Spectra of TQD-24, TQD-32, and TQD-40}

\begin{figure}[H]
\centering{}\includegraphics[scale=3]{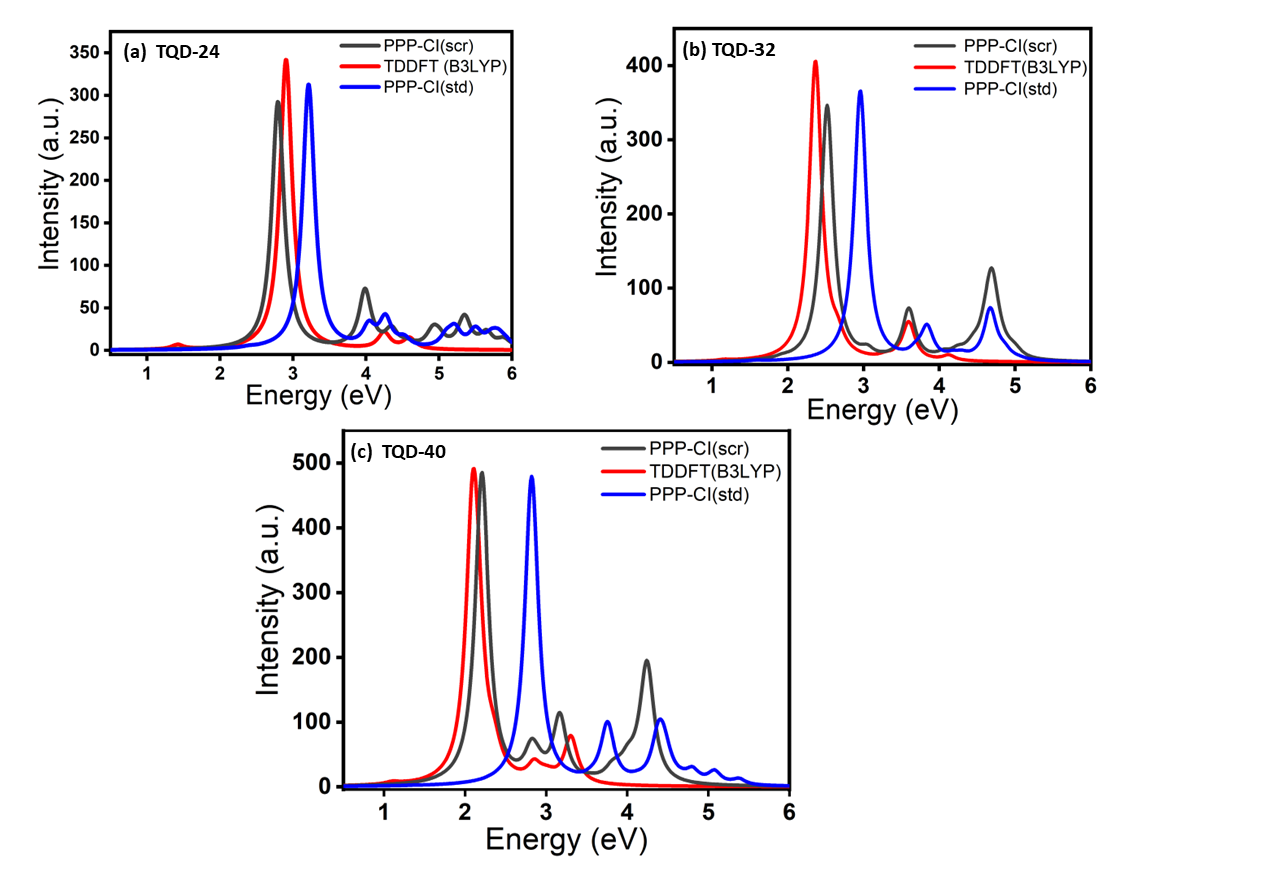}\caption{Comparison of optical absorption spectra of (a) TQD-24, (b) TQD-32
and (c) TQD-40, computed employing the PPP-CI method using both the
standard and screened Coulomb parameters with the one computed using
the first-principles TDDFT employing the B3LYP functional and $6311++G(d,p)$
basis set.}
\end{figure}

\section{PPP-CI Wave Functions of the $|\psi_{HL}\rangle$ State of Various
TQDs}

\begin{table}[H]
\centering{}%
\begin{tabular}{ccccc}
\toprule 
\multicolumn{5}{c}{$|\psi_{HL}\rangle$ state of $^{1}B_{1g}$ symmetry}\tabularnewline
\midrule
\midrule 
\multirow{2}{*}{} & \multicolumn{2}{c}{E$_{HL}$(eV)} & \multicolumn{2}{c}{Dominant configurations of the wave function}\tabularnewline
\cmidrule{2-5}
 & scr  & std  & screened  & standard\tabularnewline
\midrule 
\multirow{2}{*}{TQD-16} & \multirow{2}{*}{$2.18$} & \multirow{2}{*}{$1.80$} & $\ket{H\rightarrow L}$(0.826)  & $\ket{H\rightarrow L}$($0.811)$\tabularnewline
\cmidrule{4-5}
 &  &  & $\ket{H-1\rightarrow L+1}$(0.2437)  & $\ket{H\rightarrow L;H-1\rightarrow L+1}-c.c(0.2355)$\tabularnewline
\midrule 
\multirow{2}{*}{TQD-24} & \multirow{2}{*}{$2.11$} & \multirow{2}{*}{$1.63$} & $\ket{H\rightarrow L}(0.8148)$  & $\ket{H\rightarrow L}(0.8353)$\tabularnewline
\cmidrule{4-5}
 &  &  & $\ket{H-1\rightarrow L+1}(0.3926)$  & $\ket{H-1\rightarrow L+1}(0.1235)$\tabularnewline
\midrule 
\multirow{2}{*}{TQD-32} & \multirow{2}{*}{$1.81$} & \multirow{2}{*}{$1.66$} & $\ket{H\rightarrow L}(0.8317)$  & $\ket{H\rightarrow L}(0.6958)$\tabularnewline
\cmidrule{4-5}
 &  &  & $\ket{H-1\rightarrow L;H-2\rightarrow L}-c.c(0.1549)$  & $\ket{H-2\rightarrow L+2}(0.3965)$\tabularnewline
\midrule 
\multirow{2}{*}{TQD-40} & \multirow{2}{*}{$1.87$} & \multirow{2}{*}{$1.68$} & $\ket{H\rightarrow L}(0.8507)$  & $\ket{H\rightarrow L}(0.7103)$\tabularnewline
\cmidrule{4-5}
 &  &  & $\ket{H-2\rightarrow L+1;H\rightarrow L;H-1\rightarrow L+2}(0.0794)$  & $\ket{H-2\rightarrow L+2}(0.4189)$\tabularnewline
\bottomrule
\end{tabular}\caption{The values of the excitation energies E$_{HL}$ for TQDs calculated
using PPP-CI method using screeened (scr) and standard (std) parameters,
along with the dominant configurations contributing to their wave
functions. The coefficient of a configuration is given inside the
parentheses.}
\end{table}

\section{Spin Gaps and PPP-CI Wave Functions of the $T_{1}$ State of Various
TQDs}

\begin{table}[H]
\centering{}%
\begin{tabular}{ccccc}
\toprule 
\multicolumn{5}{c}{$T_{1}$ state of $^{3}B_{1g}$ symmetry}\tabularnewline
\midrule
\midrule 
\multirow{2}{*}{} & \multicolumn{2}{c}{E$_{ST}$(eV)} & \multicolumn{2}{c}{Dominant configurations of the wave function}\tabularnewline
\cmidrule{2-5}
 & scr  & std  & screened  & standard\tabularnewline
\midrule 
\multirow{2}{*}{TQD-16} & \multirow{2}{*}{$0.66$} & \multirow{2}{*}{$0.76$} & $\ket{H\rightarrow L}(-0.6896)$  & $\ket{H\rightarrow L}(0.6935)$\tabularnewline
\cmidrule{4-5}
 &  &  & $\ket{H-1\rightarrow L+1}$(0.3997)  & $\ket{H-1\rightarrow L+1}-c.c(0.3926)$\tabularnewline
\midrule 
\multirow{2}{*}{TQD-24} & \multirow{2}{*}{$0.75$} & \multirow{2}{*}{$0.76$} & $\ket{H\rightarrow L}(0.6233)$  & $\ket{H\rightarrow L}(0.7347)$\tabularnewline
\cmidrule{4-5}
 &  &  & $\ket{H-1\rightarrow L+1}(-0.4246)$  & $\ket{H-1\rightarrow L+1}$(0.2648) \tabularnewline
\midrule 
\multirow{2}{*}{TQD-32} & \multirow{2}{*}{$0.55$} & \multirow{2}{*}{$0.78$} & $\ket{H\rightarrow L}(0.6648)$  & $\ket{H\rightarrow L}(0.6332)$\tabularnewline
\cmidrule{4-5}
 &  &  & $\ket{H-2\rightarrow L+2}(0.3153)$  & $\ket{H-2\rightarrow L+2}(-0.4187)$\tabularnewline
\midrule 
\multirow{2}{*}{TQD-40} & \multirow{2}{*}{$0.82$} & \multirow{2}{*}{$0.90$} & $\ket{H\rightarrow L}(0.6648)$  & $\ket{H\rightarrow L}(0.6310)$\tabularnewline
\cmidrule{4-5}
 &  &  & $\ket{H-2\rightarrow L+2}(0.3649)$  & $\ket{H-2\rightarrow L+2}(0.4166)$\tabularnewline
\bottomrule
\end{tabular}\caption{Spin gap or singlet-triplet energy gap (in eV) of TGQDs computed at
the PPP-CI level using the screened (scr) and standard (std) parameters.
The dominant configurations contributing to the triplet wave functions
are also given, along with their coefficients inside the parentheses.}
\end{table}

\section{XYZ optimized coordinates of the carbon atoms employed in PPP-CI
calculations }

\begin{table}[H]
\begin{centering}
\caption{XYZ coordinates of TQD-16 used for PPP-CI calculations}
\par\end{centering}
\centering{}%
\begin{tabular}{ccc}
\toprule 
X  & \qquad{}Y  & \qquad{}Z\tabularnewline
\midrule
\midrule 
3.01955200  & \qquad{}1.73645800  & \qquad{}0\tabularnewline
\midrule 
1.64200000  & \qquad{} 1.78384000  & \qquad{}0\tabularnewline
\midrule 
0.70206700  & \qquad{}0.74270100  & \qquad{}0\tabularnewline
\midrule 
0.70206700  & \qquad{}-0.74270200  & \qquad{}0\tabularnewline
\midrule 
1.64200000  & \qquad{}-1.78384000  & \qquad{}0\tabularnewline
\midrule 
3.01955200  & \qquad{}-1.73645800  & \qquad{}0\tabularnewline
\midrule 
3.98658100  & \qquad{}-0.69090900  & \qquad{}0\tabularnewline
\midrule 
3.98658100  & \qquad{} 0.69090900  & \qquad{}0\tabularnewline
\midrule 
-1.64200400  & \qquad{}1.78384100  & \qquad{}0\tabularnewline
\midrule 
-3.01954700  & \qquad{} 1.73646100  & \qquad{}0\tabularnewline
\midrule 
-3.98658300  & \qquad{} 0.69090400  & \qquad{}0\tabularnewline
\midrule 
-3.98658300  & \qquad{}-0.69090500  & \qquad{}0\tabularnewline
\midrule 
-3.01954700  & \qquad{}-1.73646100  & \qquad{}0\tabularnewline
\midrule 
-1.64200400  & \qquad{}-1.78384100  & \qquad{}0\tabularnewline
\midrule 
-0.70206600  & \qquad{}-0.74269400  & \qquad{}0\tabularnewline
\midrule 
-0.70206600  & \qquad{} 0.74269500  & \qquad{}0\tabularnewline
\bottomrule
\end{tabular}
\end{table}

\begin{table}[H]
\begin{centering}
\caption{XYZ coordinates of TQD-24 used for PPP-CI calculations}
\par\end{centering}
\centering{}%
\begin{tabular}{ccc}
\toprule 
X  & \qquad{}Y  & \qquad{}Z\tabularnewline
\midrule
\midrule 
5.355188  & \qquad{}1.724587  & \qquad{}0\tabularnewline
\midrule 
3.943641  & \qquad{} 1.77187  & \qquad{}0\tabularnewline
\midrule 
3.025159  & \qquad{}0.756156  & \qquad{}0\tabularnewline
\midrule 
3.025159  & \qquad{}-0.756156  & \qquad{}0\tabularnewline
\midrule 
3.943641  & \qquad{}-1.77187  & \qquad{}0\tabularnewline
\midrule 
5.355188  & \qquad{}-1.724587  & \qquad{}0\tabularnewline
\midrule 
6.300784  & \qquad{}-0.707754  & \qquad{}0\tabularnewline
\midrule 
6.300784  & \qquad{} 0.707754  & \qquad{}0\tabularnewline
\midrule 
0.68412  & \qquad{} 1.808571  & \qquad{}0\tabularnewline
\midrule 
-3.943641  & \qquad{}1.771870  & \qquad{}0\tabularnewline
\midrule 
-0.68412  & \qquad{}1.808571  & \qquad{}0\tabularnewline
\midrule 
-5.355188  & \qquad{}1.724587  & \qquad{}0\tabularnewline
\midrule 
-1.602472  & \qquad{}0.728116  & \qquad{}0\tabularnewline
\midrule 
-6.300784  & \qquad{}0.707754  & \qquad{}0\tabularnewline
\midrule 
-1.602472  & \qquad{}-0.728116  & \qquad{}0\tabularnewline
\midrule 
-6.300784  & \qquad{}-0.707754  & \qquad{}0\tabularnewline
\midrule 
-0.68412  & \qquad{}-1.808571  & \qquad{}0\tabularnewline
\midrule 
-5.355188  & \qquad{}-1.724587  & \qquad{}0\tabularnewline
\midrule 
0.68412  & \qquad{}-1.808571  & \qquad{}0\tabularnewline
\midrule 
-3.943641  & \qquad{}-1.77187  & \qquad{}0\tabularnewline
\midrule 
1.602472  & \qquad{}-0.728116  & \qquad{}0\tabularnewline
\midrule 
-3.025159  & \qquad{}-0.756156  & \qquad{}0\tabularnewline
\midrule 
1.602472  & \qquad{}0.728116  & \qquad{}0\tabularnewline
\midrule 
-3.025159  & \qquad{}0.756156  & \qquad{}0\tabularnewline
\bottomrule
\end{tabular}
\end{table}

\begin{table}[H]
\begin{centering}
\caption{XYZ coordinates of TQD-32 used for PPP-CI calculations}
\par\end{centering}
\centering{}%
\begin{tabular}{ccc}
\toprule 
X  & \qquad{}Y  & \qquad{}Z\tabularnewline
\midrule
\midrule 
3.03811300  & \qquad{}-1.79816600  & \qquad{}0\tabularnewline
\midrule 
1.63113700  & \qquad{}-1.77708300  & \qquad{}0\tabularnewline
\midrule 
0.74411000  & \qquad{} -0.72194700  & \qquad{}0\tabularnewline
\midrule 
0.74410900  & \qquad{} 0.72195000  & \qquad{}0\tabularnewline
\midrule 
1.63113800  & \qquad{} 1.77708600  & \qquad{}0\tabularnewline
\midrule 
3.03811300  & \qquad{} 1.79816800  & \qquad{}0\tabularnewline
\midrule 
3.93597700  & \qquad{} 0.75158000  & \qquad{}0\tabularnewline
\midrule 
3.93597600  & \qquad{} -0.75157900  & \qquad{}0\tabularnewline
\midrule 
-1.63113600  & \qquad{} -1.77708300  & \qquad{}0\tabularnewline
\midrule 
-6.30887600  & \qquad{} -1.77909700  & \qquad{}0\tabularnewline
\midrule 
-3.03811300  & \qquad{} -1.79816600  & \qquad{}0\tabularnewline
\midrule 
-7.67164500  & \qquad{} -1.74204700  & \qquad{}0\tabularnewline
\midrule 
-3.93597600  & \qquad{} -0.75158000  & \qquad{}0\tabularnewline
\midrule 
-8.65631300  & \qquad{} -0.68315700  & \qquad{}0\tabularnewline
\midrule 
-3.93597700  & \qquad{} 0.75158100  & \qquad{}0\tabularnewline
\midrule 
-8.65631400  & \qquad{} 0.68315400  & \qquad{}0\tabularnewline
\midrule 
-3.03811300  & \qquad{} 1.79816700  & \qquad{}0\tabularnewline
\midrule 
-7.67164600  & \qquad{} 1.74204400  & \qquad{}0\tabularnewline
\midrule 
-1.63113700  & \qquad{} 1.77708500  & \qquad{}0\tabularnewline
\midrule 
-6.30887800  & \qquad{} 1.77909600  & \qquad{}0\tabularnewline
\midrule 
-0.74410800  & \qquad{} 0.72195000  & \qquad{}0\tabularnewline
\midrule 
-5.35960800  & \qquad{} 0.71989100  & \qquad{}0\tabularnewline
\midrule 
-0.74410900  & \qquad{} -0.72194800  & \qquad{}0\tabularnewline
\midrule 
-5.35960700  & \qquad{} -0.71989200  & \qquad{}0\tabularnewline
\midrule 
7.67164600  & \qquad{} 1.74204400  & \qquad{}0\tabularnewline
\midrule 
8.65631300  & \qquad{} 0.68315400  & \qquad{}0\tabularnewline
\midrule 
8.65631200  & \qquad{} -0.68315800  & \qquad{}0\tabularnewline
\midrule 
7.67164500  & \qquad{} -1.74204600  & \qquad{}0\tabularnewline
\midrule 
6.30887600  & \qquad{} -1.77909700  & \qquad{}0\tabularnewline
\midrule 
5.35960700  & \qquad{} 0.71989200  & \qquad{}0\tabularnewline
\midrule 
5.35960700  & \qquad{} -0.71989300  & \qquad{}0\tabularnewline
\midrule 
6.30887700  & \qquad{} 1.77909600  & \qquad{}0\tabularnewline
\bottomrule
\end{tabular}
\end{table}

\begin{table}[H]
\begin{centering}
\caption{XYZ coordinates of TQD-40 used for PPP-CI calculations}
\par\end{centering}
\centering{}%
\begin{tabular}{ccc}
\toprule 
X  & \qquad{}Y  & \qquad{}Z\tabularnewline
\midrule
\midrule 
-0.7156390  & \qquad{}-1.77793100  & \qquad{}0\tabularnewline
\midrule 
0.71563700  & \qquad{}-1.77793200  & \qquad{}0\tabularnewline
\midrule 
1.59744700  & \qquad{}-0.74365700  & \qquad{}0\tabularnewline
\midrule 
1.59744700  & \qquad{} 0.74365300  & \qquad{}0\tabularnewline
\midrule 
0.71564000  & \qquad{} 1.77793000  & \qquad{}0\tabularnewline
\midrule 
-0.71563700  & \qquad{} 1.77793100  & \qquad{}0\tabularnewline
\midrule 
-1.59744600  & \qquad{} 0.74365500  & \qquad{}0\tabularnewline
\midrule 
-1.59744700  & \qquad{}-0.74365400  & \qquad{}0\tabularnewline
\midrule 
3.97132800  & \qquad{}-1.78973900  & \qquad{}0\tabularnewline
\midrule 
8.61126600  & \qquad{}-1.77948500  & \qquad{}0\tabularnewline
\midrule 
5.34981500  & \qquad{}-1.80852600  & \qquad{}0\tabularnewline
\midrule 
9.99640100  & \qquad{}-1.73277000  & \qquad{}0\tabularnewline
\midrule 
6.26618800  & \qquad{}-0.74145000  & \qquad{}0\tabularnewline
\midrule 
10.95894200  & \qquad{}-0.69456100  & \qquad{}0\tabularnewline
\midrule 
6.26618800  & \qquad{} 0.74144800  & \qquad{}0\tabularnewline
\midrule 
10.95894100  & \qquad{} 0.69456600  & \qquad{}0\tabularnewline
\midrule 
5.34981400  & \qquad{} 1.80852500  & \qquad{}0\tabularnewline
\midrule 
9.99639900  & \qquad{} 1.73277400  & \qquad{}0\tabularnewline
\midrule 
3.97132800  & \qquad{} 1.78973600  & \qquad{}0\tabularnewline
\midrule 
8.61126400  & \qquad{} 1.77948600  & \qquad{}0\tabularnewline
\midrule 
3.06511900  & \qquad{} 0.71381700  & \qquad{}0\tabularnewline
\midrule 
7.67666000  & \qquad{} 0.74458500  & \qquad{}0\tabularnewline
\midrule 
3.06511900  & \qquad{}-0.71382100  & \qquad{}0\tabularnewline
\midrule 
7.67666000  & \qquad{}-0.74458700  & \qquad{}0\tabularnewline
\midrule 
-3.06512000  & \qquad{}-0.71381800  & \qquad{}0\tabularnewline
\midrule 
-3.06511900  & \qquad{} 0.71382000  & \qquad{}0\tabularnewline
\midrule 
-3.97132800  & \qquad{} 1.78973700  & \qquad{}0\tabularnewline
\midrule 
-3.97132800  & \qquad{}-1.78973600  & \qquad{}0\tabularnewline
\midrule 
-5.34981500  & \qquad{}-1.80852500  & \qquad{}0\tabularnewline
\midrule 
-6.26618800  & \qquad{}-0.74144800  & \qquad{}0\tabularnewline
\midrule 
-6.26618800  & \qquad{} 0.74145200  & \qquad{}0\tabularnewline
\midrule 
-5.34981500  & \qquad{} 1.80852700  & \qquad{}0\tabularnewline
\midrule 
-8.61126500  & \qquad{}-1.77948500  & \qquad{}0\tabularnewline
\midrule 
-9.99639900  & \qquad{}-1.73277300  & \qquad{}0\tabularnewline
\midrule 
-10.95894200  & \qquad{}-0.69456400  & \qquad{}0\tabularnewline
\midrule 
-10.95894200  & \qquad{} 0.69456200  & \qquad{}0\tabularnewline
\midrule 
-9.99640000  & \qquad{} 1.73277200  & \qquad{}0\tabularnewline
\midrule 
-8.61126700  & \qquad{} 1.77948600  & \qquad{}0\tabularnewline
\midrule 
-7.67666000  & \qquad{}-0.74458300  & \qquad{}0\tabularnewline
\midrule 
-7.67666000  & \qquad{} 0.74458600  & \qquad{}0\tabularnewline
\bottomrule
\end{tabular}
\end{table}